\documentstyle[preprint,aps]{revtex}
\tightenlines
\begin{document}
\draft
\title{Spin and density longitudinal response of quantum dots
in time-dependent local-spin-density approximation}
\author{ Ll. Serra$^1$, M. Barranco$^2$, A. Emperador$^2$, M. Pi$^2$,
and E. Lipparini$^{3}$.}
\address{$^1$Departament de F\'{\i}sica, Facultat de Ci\`encies,\\
Universitat de les Illes Balears, E-07071 Palma de Mallorca, Spain}
\address{$^2$Departament d'Estructura i Constituents de la Mat\`eria,
Facultat de F\'{\i}sica, \\
Universitat de Barcelona, E-08028 Barcelona, Spain}
\address{$^3$Dipartimento di Fisica, Universit\`a di Trento,
and INFM sezione di Trento, 38050 Povo, Italy}
\date{\today}

\maketitle

\begin{abstract}
The  longitudinal dipole response of a quantum dot has been calculated
in the far-infrared regime using local spin density functional theory.
We have studied the coupling between the  collective  spin  and
 density modes as a function of the magnetic field. We have
found that the spin dipole mode and  single particle excitations have
a sizeable overlap, and that the magnetoplasmon modes can be excited
by the dipole spin operator if the dot is spin polarized.
The frequency of the dipole spin edge mode presents an
oscillation which is clearly filling factor ($\nu$) related. We have
found that the spin dipole mode is especially soft for even $\nu$
values, becoming unstable for
magnetic fields in the region  $1 < \nu \leq 2$.
Results for selected number of electrons and confining potentials
are discussed. An analytical model which reproduces the main features
of the microscopic spectra has been developed.

\end{abstract}
\narrowtext
\section*{}

\section{Introduction}

The far-infrared (FIR) response of quantum dots is a subject
of current interest since the  experiments carried out by Sikorski and
Merkt \cite{Si89}, and by Demel et al \cite{De90}. These experiments
and the subsequent theoretical work (see Refs.
\cite{Br90,Ma90,Gu91,Sh91,Gu95,Ja98} and references therein)
have shown that the excitation spectrum of quantum dots in
the FIR region is dominated by the dipole edge magnetoplasmon peak
that
 splits into two different $B$ dispersion branches when a magnetic
field $B$ is applied perpendicularly to the dot. These peaks are
density (charge) collective modes excited by the operator
$D_{\rho}=\sum_{i=1}^N x_i$.
In the case of harmonic confinement by the potential
${1\over 2}m \omega_0^2 r^2$, as a
consequence of Kohn's theorem \cite{Ko61} the density mode is not
coupled to any other mode, and the dipole operator $D_{\rho}$
excites only two collective states at the energies
$\omega_{\pm}=\sqrt{\omega_0^2+{1\over4}\omega_c^2}\pm
{1\over2}\omega_c$, where $\omega_c$ is the cyclotron frequency.
If the confining potential is not harmonic, Kohn's theorem does not
hold, and on the one hand the energy of the modes depend on the
number
of electrons in the dot, and on the other hand a richer excitation
spectrum appears.

 Raman spectroscopy has made possible to observe in the same
sample single particle (sp), charge and spin density excitations
\cite{St94,Sc96}, whose evolution  as a function of
$B$ has been studied in  recent experiments. This
has revealed several interesting features of the sp
\cite{Lo96} and of the spin collective excitations \cite{Sc98}
in quantum dots. Limiting to the later,
the experiments have determined that the spin mode
lies  very close in energy to the
uncorrelated single electron excitations, and that  magnetoplasmons
can also be detected using spin-dependent probes. Besides, it has
been experimentally determined that the spin mode has a much lower
energy than the charge mode. These facts
constitute the body of experimental results that any theory aiming at
a quantitative simultaneous description of spin and charge
density collective modes in quantum  dots should reproduce.

The dipole spin response function for unpolarized quantum dots
at zero magnetic field has been recently addressed
by two of us \cite{Se97}. In the FIR regime, it has been found that
the response is dominated
by a low-energy collective dipole spin mode excited by the operator
$D_{m}=\sum_{i=1}^{N}{x_i\sigma^i_z}$, where $x_i$ and
$\sigma^i_z$ are Cartesian components of the
position and spin vectors, and $N$ is the number of electrons in
the dot. Similar modes have
been described in atomic nuclei \cite{Ri79},
and in alkali metal clusters \cite{Se93,Ca96}.

The aim of the present work  is to extend our previous
study to the case of a quantum dot submitted to a
perpendicular static magnetic field, which originates
a $B$ dependent spin polarization in the ground state of the dot.
We will explicitly show that this not only causes the splitting of
the spin dipole mode into two branches, one with negative and another
with positive $B$ dispersion, but also its coupling
with the dipole density mode mainly excited by the operator
$D_{\rho}$\,.
We shall see that if the confinement is not harmonic
and the dot is polarized, that operator
also excites  the dipole spin  mode.
Viceversa,  when the dot is polarized, which is  the case
if it has an odd number of electrons, or for most cases when
$B$ acts on the dot, the spin response is coupled to the density
response so that  the  external operator
$D_{m}=\sum_i{x_i\sigma^i_z}$ also excites the density
mode. When the system is fully polarized both modes coincide, while at
zero polarization they are uncoupled.

To this end, we have  self-consistently
evaluated the longitudinal response
of the dot in the time-dependent local-spin-density approximation
(TDLSDA). As longitudinal we mean  an external field which is
either spin independent, or dependent on the spin component
parallel to the magnetic field, i.e., the $z$ component.
We present results obtained for selected numbers of electrons and
confining potentials. Specifically, we have used a harmonic oscillator
potential to describe dots with $N$ = 5 and 25,
and a disk confining potential to
describe dots with  25 and 210 electrons, for which the FIR
response has been determined in detail \cite{De90}.
The results obtained for the $N$ = 5 dot have been presented as
preliminary results in Ref. \cite{Se98}. The ground state (gs)
structure of the
later two dots in intense magnetic fields has been recently addressed
\cite{Li97,Pi98}. However, no
self-consistent TDLSDA calculation for a dot as large as $N$ = 210
has been carried out before even in the density channel.

To obtain correct collective modes one needs  to have a
proper description of the ground state  these excitations are
built on. Several density functional calculations have
addressed this question \cite{Fe94,Fe95,He95,Fe97,Lu97}. The LSDA we
have used in the present work  is based on the exchange-correlation
energy functional employed in Ref.
\cite{Fe94} as an input to construct their current density
functional theory (CDFT). Within the  range of magnetic fields we
are interested in, we have checked that both LSDA and CDFT are
yielding similar results for gs properties other than the current
density. Tests of CDFT against exact and
Hartree-Fock (HF) calculations have been presented in Ref.
\cite{Fe94}. Tests of unrestricted HF against exact gs energies in
the filling factor region $2 \geq \nu \geq 1$ have also been presented
in Ref. \cite{Pa94} for small number of electrons (up to 5).

We conclude from the comparisons presented in the above references,
that TDLSA may yield fairly accurate
results for the density and spin response in the range of magnetic fields
for which experimental information is currently available.
Comparison with these experiments constitute the ultimate test of
this essentially parameter-free approximation.
Between $ 1/2 \leq \nu \leq 1$,
the results of Refs. \cite{He95,Lu97} seem to indicate that the
exchange-correlation energy of Ref. \cite{Fe94} is not very suitable
because it is based on the Levesque-Weiss-MacDonald interpolation
formula \cite{Le84}, and improved exchange-correlation energy
functionals like those of Ref. \cite{He95,Lu97} are definitely better
off (a detailed account of the accuracy of elaborated LSDA's in the
$\nu \leq 1$ regime is also given in Ref. \cite{Lu97}).
It remains to be seen whether these improvements are crucial to
describe the experimental data which for the moment are limited to
comparatively small magnetic fields.

To get a deeper understanding, simple methods  have
been developed to reproduce the gross features of the microscopic
spectra.  One such method is the sum rule
approach used in Ref. \cite{Li97} to describe multipole density
modes in quantum dots. We present here  an analytical  model, called
vibrating potential model (VPM), which  allows to understand
the TDLSDA results in a clear way, thereby providing
physical insight onto the longitudinal response of quantum dots.
The VPM model has been widely used in nuclear physics \cite{Bo75},
where it was developed  to describe nuclear collective modes. It has
also been applied to the description of simple metal clusters
 \cite{Li91}.

\section{The longitudinal response}

We consider
a  quantum dot made of $N$ electrons moving in the
$z$ = 0 plane where they are confined by the circular potential $V^+(r)$
in the presence of a constant magnetic field $B$ in the $z$ direction.
In the local-spin-density approximation (LSDA), the single electron
wave functions are given by the
solution of the Kohn-Sham (KS) equations
\begin{eqnarray}
\left[-{1\over 2}\nabla^2\right. &+& {1\over 2}\omega_c\ell_z +
{1\over 8}\omega_c^2r^2 + V^+(r) \nonumber\\
&+& \left. V^H + V^{xc} +(W^{xc}+{1\over 2}g^*\mu_B B)\sigma_z \right]
\varphi_{\alpha}(r,\theta)=\epsilon_{\alpha}\varphi_{\alpha}(r,\theta)~,
\label{eq1}
\end{eqnarray}
where
$V^H=\int{d\vec{r}\,'\rho(\vec{r}\,')/|\vec{r}-\vec{r}\,'|} $
is the Hartree potential.
$V^{xc}={\partial
{\cal E}_{xc}(\rho,m)/\partial\rho}\vert_{gs}$ and
$W^{xc}={\partial
{\cal E}_{xc}(\rho,m)/\partial m}\vert_{gs}$
are the  variations of the exchange-correlation
energy density ${\cal E}_{xc}(\rho,m)$ in the  local  approximation
taken at the ground state, and $\rho(r)$ and $m(r)$ are the
electron and spin magnetization densities. The exchange-correlation
energy density ${\cal E}_{xc}$ has been  constructed from
the results of Ref. \cite{Ta89} on the nonpolarized and fully
polarized two dimensional (2D) electron gas in the same way as in
Ref. \cite{Ko97}, i.e., using the von Barth and Hedin
\cite{Ba72} prescription to interpolate between both regimes.

We have used effective atomic units
($\hbar={e^2/\epsilon} = m = 1$), where $\epsilon$
is the dielectric constant of the semiconductor and $m$ is the electron
effective mass.
In units of the bare electron mass $m_e$ one has $m = m^*m_e$.
In this system of units,
the length unit is the effective Bohr radius $a_0^*=a_0\epsilon/m^*$,
and the energy unit is the effective Hartree
$H^*=H m^*/\epsilon^2$.
For GaAs we have taken $\epsilon$ = 12.4, $m^*$ = 0.067, and
$g^*=-0.44$, which yields
$a_0^*=97.9\, {\rm \AA}$ and $H^*\sim11.9$ meV $\sim$ 95.6 cm$^{-1}$.
In Eq.\ (\ref{eq1}) $\omega_c =e B/(m c)$ is the cyclotron
frequency
and $\mu_B={e\hbar/(2m_e c)}$ is the Bohr magneton.
The use of the same letter for the effective mass and
the spin magnetization, and for the dielectric constant and the single
electron energies should cause no confusion, since neither the
mass nor the dielectric constant will  explictly appear in the rest
of the work.

As a consequence of circular symmetry the
$\varphi_{\alpha}$'s  are eigenstates of the
orbital angular momentum $\ell_z$, i.e.,
$\varphi_{\alpha}(r,\theta)=u_{n\ell\sigma}(r)e^{-i\ell\theta}$,
with $\ell=0,\pm1,\pm2,\dots$.
The gs electron density is given by
$\rho(r)=\sum_{\alpha} n_{\alpha} |u_{\alpha}(r)|^2$,
while the gs spin magnetization density
is expressed in terms of the
spin of orbital $\alpha$, $\langle\sigma_z\rangle_\alpha$, as
$m(r)=\sum_{\alpha} n_{\alpha}
\langle\sigma_z\rangle_\alpha | u_\alpha(r)|^2$.
The numerical calculations reported in the following have  been
performed at a small but finite temperature  $T \leq$ 0.1 K, and
the KS equations have been solved by integration in $r$ space.
The thermal occupation probabilities $n_{\alpha}$ are determined by the
normalization condition
\begin{equation}
N=\sum_{\alpha} n_{\alpha} =
\sum_{\alpha}{1\over 1+{\rm exp}[(\epsilon_{\alpha}-\mu)/k_B T]}
\label{eq2}
\end{equation}
which fixes the chemical potential $\mu$.
Our iterative method works for weak and strong
magnetic fields as well, for which the effective potential is
very different. It has probed to be very robust, allowing us to
handle without any problem several hundreds of electrons.

Once the gs has been obtained, we
determine the induced densities originated by an external field
employing linear-response theory.
Following Refs. \cite{Wi83,Ra78}, we can write
the variation $\delta\rho_{\sigma}$ induced in the spin density
$\rho_{\sigma}$  ($\sigma\equiv\uparrow,\downarrow$) by an external
spin-dependent field $F$, whose non-temporal dependence we denote as
$F=\sum_{\sigma}f_{\sigma}(\vec{r})\,|\sigma\rangle\langle\sigma|$:
\begin{equation}
\delta\rho_{\sigma}(\vec{r},\omega) =
\sum_{\sigma'}\int d\vec{r}\,'\chi_{\sigma\sigma'}
(\vec{r},\vec{r}\,';\omega)f_{\sigma'}(\vec{r}\,')\; ,
\label{eq3}
\end{equation}
where
$\chi_{\sigma\sigma'}$ is the spin-density correlation function.
In this limit, the frequency $\omega$ corresponds to the
harmonic time dependence of the external field $F$ and   of
the induced $\delta\rho_\sigma$. Eq.\ (\ref{eq3}) is a 2$\times$2
matrix equation in the two-component Pauli space.
In longitudinal response theory, $F$ is diagonal in this space,
and we write its diagonal components as a vector
$F\equiv\left(\begin{array}{c} f_\uparrow\\ f_\downarrow\end{array}
\right)$. For the operators defined at the Introduction we then have

\begin{equation}
D_{\rho} \equiv   \left(\begin{array}{c} x\\ x\end{array} \right)
\,\,\,\,\,\,{\rm and}\,\,\,
D_{m} \equiv \left(\begin{array}{c} x\\ -x\end{array} \right) \,\,\, .
\label{eq4}
\end{equation}

The TDLSDA assumes that electrons respond as  free particles
to the perturbing effective field, which consists of the external
plus the induced field arising from  the changes produced
by the perturbation in the gs mean field. This
condition defines the TDLSDA correlation function
$\chi_{\sigma\sigma'}$ in terms of the
free-particle spin-density correlation function
$\chi^{(0)}_{\sigma\sigma'}$ through a
Dyson-type integral equation:
\begin{eqnarray}
\chi_{\sigma\sigma'}(\vec{r},\vec{r}\,';\omega) &=&
\chi^{(0)}_{\sigma\sigma'}(\vec{r},\vec{r}\,';\omega)\nonumber\\
&+&
\sum_{\sigma_1\sigma_2}\int d\vec{r}_1d\vec{r}_2\,
\chi^{(0)}_{\sigma\sigma_1}(\vec{r},\vec{r}_1;\omega)
K_{\sigma_1\sigma_2}(\vec{r}_1,\vec{r}_2)
\chi_{\sigma_2\sigma'}(\vec{r}_2,\vec{r}\,';\omega) \,\,\, .
\label{eq5}
\end{eqnarray}
The free particle spin-correlation function at finite
temperature  is obtained
from the KS sp wave functions, energies and occupation
probabilities:
\begin{equation}
\chi^{(0)}_{\sigma\sigma'}(\vec{r},\vec{r}\,',\omega)=
-\delta_{\sigma,\sigma'}
\sum_{\alpha\beta}\varphi^*_{\alpha}(\vec{r}\,)
\varphi_{\beta}(\vec{r}\,){n_{\alpha}-n_{\beta}\over
\epsilon_{\alpha}-\epsilon_{\beta} +\omega +i\eta}
\varphi_{\beta}^*(\vec{r}\,')
\varphi_{\alpha}(\vec{r}\,')  \,\,\, .
\label{eq6}
\end{equation}
The label $\alpha$ ($\beta$) refers to a
sp level with spin $\sigma$ ($\sigma'$) and occupation
probability $n_{\alpha}$ ($n_{\beta}$).
To simplify the analysis of the results, we have
added a small but finite imaginary part $\eta$ to the energy $\omega$.
This will make an average of the strength function by transforming
the $\delta$-peaks into Lorentzians of width $2\eta$.

The kernel $K_{\sigma\sigma'}(\vec{r},\vec{r}\,')$
is the residual two-body interaction

\begin{equation}
K_{\sigma\sigma'}(\vec{r}_1,\vec{r}_2) =
{1\over | \vec{r}_1-\vec{r}_2 |} +
\left.{\partial^2{\cal E}_{xc}(\rho,m)
\over\partial\rho_{\sigma}\partial\rho_{\sigma'}}
\right\vert_{gs}\delta(\vec{r}_1-\vec{r}_2)\; ,
\label{eq7}
\end{equation}
where
\begin{eqnarray}
\left.
{\partial^2{\cal E}_{xc}\over
\partial\rho_{\sigma}\partial\rho_{\sigma'}} \right\vert_{gs}
&=&
\left.
{\partial^2{\cal E}_{xc}\over\partial\rho^2}
\right\vert_{gs} +
(\eta_{\sigma}+\eta_{\sigma'})\,
\left.{\partial^2{\cal E}_{xc}\over\partial\rho\,\partial
m}\right\vert_{gs} +
\eta_{\sigma}\eta_{\sigma'}\,
\left.{\partial^2{\cal E}_{xc}\over\partial
m^2}\right\vert_{gs} \nonumber\\
&\equiv&
K(r) +
(\eta_{\sigma}+\eta_{\sigma'})\, L(r) +
\eta_{\sigma}\eta_{\sigma'}\; I(r) \;,
\label{eq8}
\end{eqnarray}
with $\eta_{\uparrow}=1, \eta_{\downarrow}=-1$. The last expression
is the definition of the $K$, $L$, and $I$ functions.

When the system is not polarized, there are only two independent
correlation
functions. These are $\chi_{\rho\rho}$ and $\chi_{mm}$ describing,
respectively,
the density response to $D_{\rho}$ and the spin
response to $D_{m}$. They are given by
\begin{eqnarray}
\chi_{\rho\rho}&=&\chi_{\uparrow\uparrow}+\chi_{\downarrow\downarrow}+
\chi_{\uparrow\downarrow}+\chi_{\downarrow\uparrow} \nonumber\\
\chi_{mm}&=&\chi_{\uparrow\uparrow}+\chi_{\downarrow\downarrow}-
\chi_{\uparrow\downarrow}-\chi_{\downarrow\uparrow}  \,\,\,   ,
\label{eq9}
\end{eqnarray}
and the four equations (\ref{eq5}) reduce to two uncoupled equations
for $\chi_{\rho\rho}$ and $\chi_{mm}$ whose kernels are
given by $1/r_{12} + K \delta(r_{12})$ and $I\delta(r_{12})$,
respectively,  and the free-particle correlation
function $\chi^{(0)}=\chi^{(0)}_{\uparrow\uparrow}+
\chi^{(0)}_{\downarrow\downarrow}=
2\chi^{(0)}_{\uparrow\uparrow}$ is the same in both channels because
$\chi^{(0)}_{\uparrow\uparrow} = \chi^{(0)}_{\downarrow\downarrow}$.
This constitutes the paramagnetic limit of the longitudinal response
with uncoupled density and spin channels\cite{Se97}, in which the
residual interaction consists of a Coulomb direct plus an
exchange-correlation terms in one case,
and only of an exchange-correlation term in the other.

When the system is polarized one no longer has
$\chi^{(0)}_{\uparrow\uparrow} = \chi^{(0)}_{\downarrow\downarrow}$,
and there are two more independent correlation functions
\begin{eqnarray}
\chi_{\rho m}&=&\chi_{\uparrow\uparrow}-\chi_{\downarrow\downarrow}-
\chi_{\uparrow\downarrow}+\chi_{\downarrow\uparrow} \nonumber\\
\chi_{m \rho}&=&\chi_{\uparrow\uparrow}-\chi_{\downarrow\downarrow}+
\chi_{\uparrow\downarrow}-\chi_{\downarrow\uparrow}\; ,
\label{eq10}
\end{eqnarray}
which produce the density response to  $D_{m}$ and the
spin response to $D_{\rho}$, respectively.

Equations (\ref{eq5}) have been solved as a generalized matrix
equation in coordinate space after
performing an angular decomposition of  $\chi_{\sigma\sigma'}$ and
$K_{\sigma\sigma'}$ of the kind
\begin{equation}
K_{\sigma\sigma'}(\vec{r},\vec{r}\,')=
\sum_{\ell}K^{(\ell)}_{\sigma\sigma'}(r,r') e^{i \ell(\theta -\theta')}
\; .
\label{eq11}
\end{equation}
Only modes with $\ell=\pm 1$ couple
to the external  dipole fields $D_\rho$ and $D_m$.
This can be readily seen performing the
angular integral in Eq. (\ref{eq3}). In practice, we have
considered the
multipole expansion of the external field, using the
dipole vectors
\begin{equation}
D_\rho^{(\pm 1)} = \frac{1}{2} r e^{\pm i\theta}
\left(\begin{array}{c} 1\\ 1\end{array}\right)
\,\,\, {\rm and} \,\,\,
 D_m^{(\pm 1)}   =   \frac{1}{2} r e^{\pm i\theta}
\left(\begin{array}{c} 1\\ -1\end{array}\right) \,\,\, .
\label{eq12}
\end{equation}
For a polarized system having a non zero magnetization in the gs,
 the $\ell=\pm 1$ modes are not
degenerate and give rise to two excitation branches with
$\Delta L_z=\pm 1$, where $L_z$ is the gs orbital
angular momentum.

The response functions corresponding to the above dipole fields
have been obtained from the
$\ell=\pm1$ components of the correlation functions
$\chi_{AB}^{(\pm 1)}(r,r';\omega)$ with $A, B =\rho, m$ as:
\begin{eqnarray}
\alpha_{AB}(\omega)&=& \pi^2\int d r_1\, d r_2 \,r_1^2\, r_2^2\,
(\chi_{AB}^{(+1)}(r_1,r_2;\omega)+
\chi_{AB}^{(-1)}(r_1,r_2;\omega)) \nonumber\\
&\equiv&
\alpha_{AB}^{(+1)}(\omega) + \alpha_{AB}^{(-1)}(\omega)
\label{eq13}
\end{eqnarray}
Their imaginary parts are related to the
strength functions as $S_{AB}(\omega)=
{1\over\pi} {\rm Im}[\alpha_{AB}(\omega)]$.
Although the excitation energy $\omega$ and strength functions are
always  positive, it may be easily verifed that the following
relations formally hold:
\begin{eqnarray}
{\rm Re}\left[\alpha_{AB}^{(-\ell)}(\omega) \right]
&=& {\rm Re}\left[\alpha_{AB}^{(\ell)}(-\omega)\right] \nonumber\\
{\rm Im}\left[\alpha_{AB}^{(-\ell)}(\omega) \right]
&=&-{\rm Im}\left[\alpha_{AB}^{(\ell)}(-\omega)\right] \; .
\label{eq14}
\end{eqnarray}

To check the numerical accuracy of the calculations we have used
the f-sum rules for the dipole operators, which can be
expressed in terms of gs quantities \cite{Li89}:
\begin{eqnarray}
m^{(\rho\rho)}_1 &=& \int S_{\rho\rho}(\omega)\,\omega\, d\omega
={1\over2}\langle 0|[D_{\rho},[H,D_{\rho}]] | 0\rangle ={N\over2}
\nonumber\\ m^{(mm)}_1 &=& \int S_{mm}(\omega)\,\omega\, d\omega
={1\over2}\langle 0|[D_{m},[H,D_{m}]] | 0\rangle={N\over2}
\nonumber \\
m^{(m\rho)}_1 &=&
m^{(\rho m)}_1 = \int S_{m\rho}(\omega)\,\omega\, d\omega+
\int S_{\rho m}(\omega)\, \omega \,d\omega
=\langle 0 |[D_{m},[H,D_{\rho}]] | 0 \rangle= 2S_z ~~
\,\, ,
\label{eq15}
\end{eqnarray}
where $S_z$ is the total spin of the ground state.

\section{Results}

Figures \ref{fig1}-\ref{fig3} display the dipole  strength function of
the $N$ = 5, 25, and 210 dot for selected  $B$ values.
Solid lines correspond to the  density response to $D_{\rho}$,
and dotted lines to the spin response to $D_{m}$, that is to
$S_{\rho\rho}$ and to $S_{mm}$. Dashed lines represent the
free particle strength function. For the five
electron dot we have used a parabolic confining potential
$V^+(r)={1\over 2}m \omega_0^2 r^2$ with $\omega_0$ = 4.28  meV, and
for the other dots we have used a disk confining potential as
described in Refs. \cite{Li97,Pi98}.

For $N$ = 25 and 210, most $B$ values displayed correspond to integer
filling factors $\nu$. It was found in Ref. \cite{Pi98} that
for $N$ = 210 and an $R$ disk confining potential,
these values follow the law $\nu = 2 \pi c\, n_s/(e B)$
pertaining to the 2D system, where $n_s = N/(\pi R^2)$ is the electron
surface density.
 For the $N$ = 25 dot, that law yields $B$ = 3.29 T
as the value at which the system becomes fully polarized.
Actually, we have found that the dot is in the maximum density drop
(MDD) state for 3.42 T $\le B \le$ 3.70 T.
We have taken as  $\nu$ = 1  configuration that corresponding
to $B_1$ = 3.56 T, and have defined the
other $\nu$ configurations  as those corresponding to the
value  $B_{\nu} = B_1/\nu$.

The results of Ref. \cite{De90} for the $N$ = 25 dot seem to indicate
that a confining potential of parabolic type might be more adequate
(see the discussion of the charge mode at the end of this Section).
Consequently,  we have also studied it using a
harmonic confining potential with
$\omega_0$ = 2.78 meV to reproduce the experimental dipole
energy at $B$ = 0. In this case
 the system is in the MDD state for 4.46 T $\le B \le$ 4.69 T.
We have taken $B_1$ = 4.58 T and have defined the  other
$\nu$ configurations  as indicated before.

The existence of a MDD state for this dot is in agreement with the
findings of Ref. \cite{Fe97}. It is also worth to remark that
even for such a small dot, $\nu$ as defined before
also coincides with the number of
occupied $(n,\uparrow {\rm or} \downarrow)$ bands for values up
to $\nu$ = 5-6, which correspond to rather low $B$ intensities.
We present in Fig. \ref{fig4} the strength function corresponding
to $N$ = 25 with parabolic confinement.

Figures \ref{fig1}-\ref{fig4} show that in both channels the
response at $B$ = 0 is concentrated within a small energy range, with
one single peak or with several closely lying fragmented peaks which
exhaust most of the f-sum rule. The peak energy is lower in the spin
than in the density channel. This is due to
the character of the residual interaction, which is attractive in the
former and repulsive in the later channel, and shifts the
TDLSDA responses from the free particle response in opposite
directions.
The residual interaction in  the spin channel is weaker than in
the density channel, where not only the exchange-correlation
but also the  Coulomb direct term contributes.
Consequently, the spin response is  close in energy to the free
response.  It is thus difficult to distinguish
the collective spin mode from the single particle spectrum. In large
dots it also causes a stronger Landau damping  in the spin than
in the density channel.
These facts have been observed and discussed in Ref. \cite{Sc98}.

At $B$ = 0, as a consequence of Kohn's theorem,
if the confining potential is harmonic with frequency $\omega_0$
the excitation energy of the dipole density mode is equal to
$\omega_0$ irrespective of the number of electrons. Otherwise, the
excitation energy depends on $N$, see for example
Refs. \onlinecite{Gu91,Se97}.
In the spin channel Kohn's theorem does not hold, and a
size dependence appears in the dipole spin mode even for parabolic
confining potentials \cite{Se97}.

When $B$ is not zero, the dipole mode in either channel  splits
into two branches, one with negative and another with positive
$B$ dispersion. The splitting is
due to the breaking of the $\ell$-degeneracy of the sp
energies by the applied magnetic
field. Several new phenomena then appear.
We first notice that for $B$ values such that the  spin of the
 dot gs is different from zero,
the spin and density modes are coupled. This is particularly apparent
in the $N$ = 210 dot. Indeed, at $B$ = 1.71 and 5.14 T
 the system is almost paramagnetic, having 2$S_z$ = 2 and 0,
respectively  (see Fig. 4 of Ref.  \cite{Pi98}).
As a consequence,  the modes are uncoupled, as it can be seen from
panel (b) and especially from panel (d) of Fig. \ref{fig3}. In
contradistinction, at $B$ =  3.43 and 7 T we have 2$S_z$ = 54 and
74: the system has a large spin magnetization in
the gs and the spin and density modes are clearly coupled, as displayed in
panels (c) and (e) of that figure. One can
see a distinct peak in the spin response at the energy of the
density mode. This effect has been experimentaly observed
 \cite{Sc98}. The strength
of this peak increases with $S_z$ and when the system is fully
polarized, which happens sligthly above $B$ = 10 T for the $N$ = 210
dot,
all the strength is transferred from the spin to the density channel.
Conversely, the spin mode may be observed in the density
channel with some intensity.
This effect is hindered because  Kohn's theorem prevents it from
occuring for parabolic potentials, and for the disk potential
it is of order $(2 S_z/N)^2$.

The interplay between charge and spin modes is especially marked
 in the mixed channel where
the density response to the spin dipole operator $D_m$, or the spin
response to the density dipole operator $D_{\rho}$ are described.
This is  shown in Fig. \ref{fig5} for the N=210 dot at $\nu$ = 3.
 One clearly observes two peaks  at the energy of the
density modes,  and another two at the energy of the spin modes.
It can be understood
casting the mixed response into a sum over 'spin dipole states'
$| m \rangle$ and another over 'charge dipole states' $| \rho \rangle$
\begin{eqnarray}
S_{m\rho}(\omega) = S_{\rho m}(\omega)&=&
\sum_n \langle 0|D_{\rho}| n \rangle
\langle n|D_m | 0 \rangle  \delta(\omega - \omega_{n0})
\nonumber\\
&=& \sum_{\rho} \langle 0|D_{\rho}| \rho \rangle
\langle \rho|D_m | 0 \rangle  \delta(\omega - \omega_{\rho0})
\nonumber\\
&+& \sum_m \langle 0|D_{\rho}| m \rangle
\langle m|D_m | 0 \rangle  \delta(\omega - \omega_{m0})
\,\, .
\label{eq15b}
\end{eqnarray}
For a disk confining potential, the matrix element
$\langle 0|D_{\rho}| m \rangle$ is not zero and there is a contribution
to $S_{m\rho}$ from the spin modes. For a harmonic confining potential
$\langle 0|D_{\rho}| m \rangle$ is  zero and
only the density modes would contribute to $S_{m\rho}$ through the
$\rho$-sum in Eq. (\ref{eq15b}).

The $B$ dispersion of the main peaks of the spectrum is reported
in Figs. \ref{fig6}$-$\ref{fig9}. In these figures the density modes
are represented by dots and the spin modes by triangles. Solid symbols
correspond to integer filling factor values, and the inserts
show the negative $B$ dispersion branch of the spin mode. As a guide,
we have drawn lines starting at the value of the  $B$ = 0 frequency
and following the VPM $B$ dispersion laws (see Eqs. \ref{eq23}).

Several features of these figures are worth to  discuss.
Concerning the spin modes, we first see that at low $B$
their energy is much smaller than the energy of the density
modes, in agreement with the
experimental findings of Ref. \cite{Sc98}.
At higher $B$ the dot is eventually fully polarized and the
longitudinal spin and density modes merge, as in the two dimensional
electron gas (see Fig. 5 of Ref. \cite{Ka84}).
This is not explicitly shown in the figures. Secondly, the negative
$B$ dispersion branch of the spin mode manifests a clear oscillatory
behaviour with $\nu$, similar to that found for the
density response \cite{Bo96}, also discussed in Ref. \cite{Pi98}:
the 'paramagnetic' even $\nu$ configurations have softer spin modes
than the 'ferromagnetic' odd $\nu$ configurations.

Our calculation predicts a spin instability occurring when the
energy of the spin mode lower branch
goes to zero at a critical $B$ between $\nu$ = 1
and 2. This instability also manifests in the static spin
polarizability ${\rm Re}[\alpha_{mm}^{(-1)}(0)]$, which becomes negative
at these large $B$'s. This indicates that the gs is no longer an
energy minumum and thus is not stable. It is worth to recall that
no collective spin dipole modes are observed in the experiments
at these rather high  $B$ values.
However, one cannot discard this might
be due to the strong Landau damping existing in this energy region.
We remark that within LSDA, a spin wave instability of similar type
has been found in parabolically confined dots at $B$ = 0
\cite{Ko97} whose origin is a spontaneous breaking of the circular
symmetry, which could also be the origin of the spin dipole instability.

It is of some interest
to look at the effect that a moderate temperature
may have on the spin response of the $N$ = 210 dot at high magnetic
fields. As an example, we
have considered $B$ = 5.14 T ($\nu$ = 2), and  $B$ = 7
T, and have obtained the spin and charge responses at $T$ = 2 K.

The results are displayed in Fig. \ref{fig9b}. Apart from the well
known $T$-independence of the charge response at low temperatures, the
figure shows that the low energy spin peak recovers at $T$ = 2 K, no
longer lying at zero energy \cite{note}. We do not display the results
for $\nu$ = 2, since  the response
remais essentially unchanged. The different
behavior is due to the quite distinct sp spectrum of
these configurations, as can be seen from Figs. 2(b) and 5(d) of Ref.
\cite{Pi98}, which makes the $B$ = 7 T state very sensitive to thermal
effects (see Fig. 7 of Ref. \cite{Pi98}). It is a general rule that
integer $\nu$ states are rather inert thermally because for them
the chemical potential lies between Landau bands and consequently,
very few sp states are affected by a change in $T$ if it is much
smaller than the sp energy gap, which is the present case. The
situation is completely different for $B$ values that do not
correspond to integer $\nu$ values. In this situation, the chemical
potential is on top of a Landau band, and the changes are sizeable
because of the large density of states around the chemical potential
(see Figs. 2, 5 and 7 of Ref. \cite{Pi98}).

Finally, we would like to comment on  the
density dipole mode. For the parabolic confining potential, Figs.
\ref{fig6} and \ref{fig9} show the well-known result that the density
response yields the classical law represented by the first of
Eqs. \ref{eq23}. For the disk confining potential, that law is also
fairly obeyed, particularly by the negative $B$ dispersion branch.
We have systematically found that the positive $B$ dispersion
branch is fragmented, especially for the $N$ = 210 dot. Comparing our
calculations for the $N$ = 25 dot
with the experimental results \cite{De90}, we conclude that a parabolic
potential is better suited than a disk potential
to represent the physical situation. On the contrary,
the confining potential of the $N$ = 210 dot is not parabolic,
and this is the origin of the second upper branch found in the
experiment.
The non-harmonicity  of the confining potential has been presented
in Ref. \cite{Gu91} as the origin of that branch on the basis of a
Hartree$+$random phase approximation (RPA) calculation in dots with
$N
\leq$ 30 (see also Ref. \cite{Ye94}). Our TDLSDA calculation
supports that interpretation.

It is worth to see that the density response of
such a large dot displayes two  instead of
one satellite branch  (compare with the results for the small $N$ =
25 dot), i.e.,  TDLSDA seems to produce
a high frequency peak that is more fragmented than in the experiments.
We want to remark that the upper branches disappear at intense $B$,
in agreement with the
experimental findings. This gives further support to our explanation
and that of \cite{Gu91} about the origin of these branches.
Indeed, at higher magnetic fields one expects
that the harmonic cyclotron potential dominates over the other
contributions in the KS equation.

\section{THE VIBRATING POTENTIAL MODEL}

The intuitive idea behind TDLSDA is that a small amplitude
time-dependent
variation of the mean field around the static equilibrium
configuration produces an oscillation in the electronic density,
which causes a small amplitude collective motion of the
system. This motion is  self sustained
if the induced density is   precisely
that needed to generate the oscillating potential. The vibrating
potential model naturally arises when one considers the first
iteration of the \mbox{(perturbed mean field)} $\longleftrightarrow$
\mbox{(induced density)} selfconsistent scheme. In homogenous systems,
where translational invariance determines the shape of the induced
density, it yields the exact solution. In finite systems
(nuclei, metal clusters, dots), the model provides a useful
approximation.

Using the general method described in the Appendix,
we consider the following  VPM Hamiltonian
\begin{eqnarray}
H&=&\sum_{i=1}^N\left[-{\nabla^2\over2} + {1\over2}\omega_c\ell_z
+ {1\over 8}\omega_c^2r^2
+ {1\over 2}\omega_{xc}^2r^2+({1\over2}g^*\mu_B B
+ {V_{\sigma}\over \rho_0}m_0)\sigma_z
+ \delta V(\vec{r}\, ,t)\right]_i   \nonumber \\
&=&\sum_{i=1}^N\,\left[ H_0(\vec{r}_i) + \delta V(\vec{r}_i\,
,t)\right]
\label{eq16}
\end{eqnarray}
with a time-oscillating potential
\begin{equation}
\delta V(\vec{r}\,,t)=-{1\over N}
\left[(\omega_{xc}^2+\omega_0^2)\langle\sum_k x_k\rangle x
-2 {V_{\sigma}\over\langle r^2\rangle}
\langle\sum_k x_k\sigma_z^k\rangle x\sigma_z \right] ~~,
\label{eq17}
\end{equation}
where $\omega_0^2=\pi\,n_s/R$.
The time dependence of $\delta V(\vec{r}\, ,t)$ is
in the $\langle...\rangle$ spatial foldings with the densities
induced by a time dependent external field, see the Appendix.
To get the static part $H_0$ of the
Hamiltonian Eq. (\ref{eq16}) we have assumed an exact cancellation between
the Hartree and the  external potentials, and have taken a parabolic
approximation for the exchange-correlation potential at $B=0$.
$V_{\sigma}$ is the exchange-correlation constant introduced in the
Appendix.

This VPM Hamiltonian can now be solved analytically within RPA
by finding the operators $O^+$ solution
of the equations of motion
\begin{equation}
[H,O^+]=\omega \,O^+ ~~.
\label{eq18}
\end{equation}
We have used the methods illustrated in Ref.\onlinecite{Li89} to compute
the conmutators with the Hamiltonian as
\begin{equation}
[H,O]=[H_0,O] + \delta V(O) ~~,
\label{eq19}
\end{equation}
where $H_0$ is the static Hamiltonian and $\delta V(O)$ the
variation arising from the induced densitites.
It can be shown that the solutions to Eq. (\ref{eq18}) are
given by
\begin{eqnarray}
O^{\rho\,+}_{\pm}&=& {1\over2}\sqrt{{\bar\omega\over N}}
\left(Q_{\pm}-{i\over \bar\omega}P_{\pm}\right)  \nonumber\\
O^{m\,+}_{\pm}&=&{1\over2}\sqrt{{\bar\omega_\sigma\over
N[1-({2S_z\over N})^2]}}
\left[ (Q^{\sigma}_{\pm}-{i\over \bar\omega_{\sigma}}P^{\sigma}_{\pm})
-{2S_z\over N}(Q_{\pm}-{i\over \bar\omega_{\sigma}}P_{\pm}) \right]
\,\, , \label{eq20}
\end{eqnarray}
where
\begin{eqnarray}
Q_{\pm}=\sum_{i=1}^N(x_i \pm y_i) ~~&,&
~ P_{\pm}=\sum_{i=1}^N(p_{xi} \pm p_{yi}) \nonumber\\
Q^{\sigma}_{\pm}=\sum_{i=1}^N(x_i \pm y_i)\sigma_z^i ~~&,&~~
P^{\sigma}_{\pm}=\sum_{i=1}^N(p_{xi} \pm p_{yi})\sigma_z^i
\label{eq21}
\end{eqnarray}
and
\begin{equation}
\bar\omega=\sqrt{\omega_o^2+{\omega_c^2\over4}}~~,~~\bar\omega_{\sigma}=
\sqrt{\omega_{xc}^2+{2V_{\sigma}\over\langle
r^2 \rangle}+{\omega_c^2\over4}}~~.
\label{eq22}
\end{equation}
The corresponding frequencies are
\begin{eqnarray}
\omega^{\rho}_{\pm}&=&\bar\omega\pm{\omega_c\over2}  \nonumber\\
\omega^m_{\pm}&=&\bar\omega_{\sigma}\pm{\omega_c\over2}~~,
\label{eq23}
\end{eqnarray}
and it is easy to verify that
\begin{equation}
[L_z,O^+_{\pm}]=\pm \,O^+_{\pm}~~.
\label{eq24}
\end{equation}
The states $| \omega^{\rho\,,m}_{\pm}\rangle \equiv O^{\rho\,,
m\,+}_{\pm} |0\rangle$ are  orthonormal and
carry an orbital angular momentum $L_0\pm1$ and a spin $S_0$,
where $L_0$
and $S_0$ are the orbital and spin angular momenta of the ground state,
respectively.

The charge dipole and spin dipole strengths are distributed among the above
states as follows:
\begin{eqnarray}
| \langle0 | D_{\rho}| \omega_+^{\rho}\rangle |^2&=&
|\langle 0 | D_{\rho}|\omega_-^{\rho}\rangle|^2=
{1\over4}{N\over\bar\omega}
\nonumber
\\
|\langle0| D_m |\omega_+^{\rho}\rangle|^2&=&
| \langle0|  D_m| \omega_-^{\rho}\rangle|^2=
{S_z^2\over\bar\omega N}
\nonumber
\\
| \langle0|  D_{\rho}| \omega^m_+\rangle|^2&=&
| \langle0|  D_{\rho}| \omega^m_- \rangle |^2= 0
\nonumber
\\
|\langle0|  D_m| \omega^m_+\rangle |^2&=&
| \langle0|  D_m| \omega^m_-\rangle |^2=
{1\over4}{N\over\bar\omega_{\sigma}}
\left[1-\left({2 S_z\over N}\right)^2\right]~~.
\label{eq25}
\end{eqnarray}
It is a simple matter to check that the above solutions
exhaust the sum rules Eq. (\ref{eq15}).

This vibrating potential model  reproduces
the most salient features of the full self-consistent
calculation. Its parabolic form guarantees that  Kohn's theorem is
fulfilled, and as a consequence  the response to
$D_{\rho}$ is shared by just two peaks, which
have the same strenght $N/4\bar\omega$. Accordingly, the
spin dipole modes
$|\omega^m_{\pm}\rangle$ are not excited by the dipole
operator $D_{\rho}$ and the corresponding matrix elements vanish.
Another consequence of Kohn's theorem is that within VPM only
the density mode contributes to the mixed response.
The model predicts that in the mixed channel, the charge dipole modes
$|\omega^{\rho}_{\pm}\rangle$ are excited with the same strenght
$S_z^2/\bar{\omega} N$
by the spin dipole operator  $D_m$.

In the spin channel, the spin dipole operator  $D_m$ excite both the
charge
$|\omega^{\rho}_{\pm}\rangle$ and spin $|\omega^m_{\pm}\rangle$
modes.  These peaks have  strengths
$|\langle 0 | D_m |\omega^m_{\pm}\rangle |^2 =
N [1-(2 S_z/N)^2]/4\bar\omega_{\sigma}$
and $|\langle0| D_m|\omega^m_{\pm}\rangle|^2$ =
$S_z^2/\bar\omega N$ , which are the same for both
$\Delta L_z=\pm1$ branches.
Finally, when $S_z=0$
the density and spin modes  are decoupled, and when the system
is fully polarized, i.e., $2S_z=N$, all the strength is transferred
to the density channel.

Besides this qualitative agreement, the strengths
given by the first and last lines of Eq.\ (\ref{eq25}) agree
well with the result of the microscopic calculation.
Also the ratio $(2S_z/N)^2$ between  the strength of the
density peaks excited by $D_m$ and $D_{\rho}$ is reproduced.

The second Eq. \ (\ref{eq23}) can be used to determine the energy
of the spin dipole mode  $\omega^m_{\pm}$ as the first one
has been often used in the case of the charge dipole
mode, if we fix the parameters entering that equation.
We first take
$\omega_{xc}^2=2\pi n_s/R^2$. This estimate is obtained identifying
the kinetic energy per particle  in the exchange-correlation
harmonic oscillator potential  with that of the 2D Fermi
gas $\pi n_s /2$,
and approximating the mean square radius of the dot
$\langle r^2\rangle$ by $R^2/2 =r_s^2 N/2$, where $r_s^2 = 1/(\pi n_s)$.
 One then gets
\begin{equation}
\omega^m_{\pm}=
\sqrt{{4\over r_s^4 N} (r_s^2 V_{\sigma} +{1\over2})+
{\omega_c^2\over4}} \pm{\omega_c\over2}~~.
\label{eq26}
\end{equation}

The value of $r_s^2 V_{\sigma}$ is related to the
spin susceptibility of the two-dimensional electron gas
$\chi_0/\chi = r_s^2\,V_{\sigma} + 1$ (Ref. \cite{Ta89}).
Eq. (\ref{eq26}) yields values for $\omega^m_{\pm}$
which agree with the TDLSDA ones.
In particular, we have checked that the $N$ dependence of the $B$ = 0
calculations reported in Ref. \cite{Se97}
 is well reproduced taking  $r_s^2 V_{\sigma} \simeq -0.3$ at
 $r_s\simeq 1$.

 Notice that the $N$ and $r_s$ dependence of energy of the spin mode
is different from that of  the density mode, which is  given by
\begin{equation}
\omega^{\rho}_{\pm}=
\sqrt{{1\over r_s^3 N^{1/2}}+{\omega_c^2\over4}} \pm{\omega_c\over2}
\label{eq27}
\end{equation}
when we take for the frequency $\omega_0$ of the confining potential
the estimate $\omega_0=r_s^{-3/2}N^{-1/4}$ obtained from the disk
potential in the $r/R<<1$ limit.

We report in Fig. \ref{fig10} the energies given by Eq. (\ref{eq26})
for a dot of  $N$ = 200 together with the
experimental  spin dipole mode of Ref. \cite{Sc98}.
A value of $r_s = 0.65$  has been used which yields
$r_s^2 V_{\sigma} = -0.24$. From the figure one sees that
the negative $B$ dispersion experimental branch is fairly well
reproduced. It has been experimentaly found that the dipole positive
$B$ dispersion branch, as well as other
 spin modes laying at higher energies are rather  $B$ independent,
so one should not expect that the simple model leading to Eq.
(\ref{eq26}) reproduces this behavior.

The VPM works better for large than for small dots. This is not
surprising in view of the approximations leading to VPM. It can be
seen comparing with the TDLSDA results in Figs. \ref{fig6}-\ref{fig9}
after being advised that for the $N$ = 5 dot, the dotted line
representing $|\omega_+^m\rangle$ is not passing near the
main peak energies of the positive $B$ dispersion branch,
but near to the satellite peak energies.

Eq. (\ref{eq26}) shows that within VPM, the spin instability at
$B$ = 0 mentioned in the previous Section occurs when
$r_s^2\,V_{\sigma} + 1/2 \leq 0$. In the  2D electron gas
it happens when $r_s^2\,V_{\sigma} + 1 \leq 0$ \cite{Ta89}. The
difference between these conditions is due to finite size
effects, which are crucial to determine the values of $r_s$
at which the instability appears in quantum dots. Whereas in the bulk
the spin instability sets in at $r_s \simeq 37$,
Eq. (\ref{eq26}) yields that in dots it happens at $r_s \sim 3$. This
value is well within the range of those found in Ref. \cite{Ko97}
for the onset of a spin density wave instability in small magic dots,
and a factor of two larger than the typical values obtained in that
reference for open shell dots.

\section{Summary and outlook}

In this work we have thoroughly discussed the longitudinal dipole
response of
quantum dots. We have shown that TDLSDA is able to reproduce the
main features of the experimental results. In particular, we have
found that the density and spin modes are clearly coupled in the spin
channel if the system is partially polarized,
and that the spin modes are especially soft for even filling
factors. We have determined that at $B \neq 0$ the spin response lies
in the energy region of single electron excitations.

We predict that the frequency of the spin edge
magnetoplasmon presents an oscillatory behavior as a function of $\nu$,
and that the spin dipole mode becomes unstable at $B$ values  in the
$1 < \nu \leq 2$ region.

Our numerical scheme has allowed us to study large size dots
 whose spectrum has been experimentally studied in detail,
instead of relying on  extrapolation of the results obtained for small
size dots. This is crucial to identify the $\nu$ behavior of
physical quantities like excitation frequencies.

TDLSDA can be easily applied to other multipole spin and density
excitations. This is relevant in view that  recent experiments
have been able to identify the spin monopole and
quadrupole modes, and likely charge modes different form dipole
 \cite{Sc98}.
Work to extend the present study to other multipolarities, and to
study the spin transverse channel in large dots along the line
described in Ref. \cite{Li98} is under way.

\acknowledgements

This work has been performed under grants PB95-1249 and PB95-0492
from CICYT, Spain, and GRQ94-1022 from
Generalitat of Catalunya. A.E. acknowledges support from the
Direcci\'on General de Ense\~nanza Superior (Spain).

\appendix

\section*{}

In this Appendix we provide a general derivation of the VPM
Hamiltonian leading to Eq.\ (\ref{eq16}).
We start from the density and magnetization variations
\begin{eqnarray}
(\delta\rho(\vec{r},t), \delta m(\vec{r},t) ) &=&
\alpha(t) ( \nabla_x\rho_0(\vec{r}\,), \nabla_x m_0(\vec{r}\,) )
\nonumber\\
&+& \beta(t) ( \nabla_x m_0(\vec{r}\,), \nabla_x\rho_0(\vec{r}\,) )
~~,
\label{aa1}
\end{eqnarray}
 obtained from the static  ground state density
$\rho_0$ and spin
magnetization $m_0$ through the unitary transformations
$e^{\alpha(t)\sum_i\nabla_x^i}$ and
$e^{\beta(t)\sum_i\nabla_x^i\sigma_z^i}$, respectively, in the limit
of small deformations. In Eq.\ (\ref{aa1}) $\alpha(t)$ and
$\beta(t)$ give the amplitude of the oscillations.

Assuming for the exchange-correlation potentials the forms
$V^{xc}=V^{xc}(\rho)$  and $W^{xc}= m V_{\sigma} / \rho_0$,
which amounts to expand ${\cal E}(\rho, m)$ around $m$ = 0 up to $m^2$
order and identifying $V_{\sigma}$ with $\rho_0 \partial^2 {\cal
E}_{xc} (\rho_0, m)/\partial m^2|_{m=0}$,
 one obtains the variation   in the
one body potential of Eq.\ (\ref{eq1}) induced by the density
variations
(\ref{aa1}):
\begin{equation}
\delta V(\vec{r}_i,t) = \alpha(t)\left(Q(\vec{r}_i) +
{2S_z\over N}Q_{\sigma}(\vec{r}_i)\,\sigma_z^i\right) +
\beta(t)\left({2S_z\over N}Q(\vec{r}_i) +
Q_{\sigma}(\vec{r}_i)\,\sigma_z^i\right) ~~,
\label{aa2}
\end{equation}
where
\begin{equation}
Q(\vec{r}\,)=\left(\nabla_x V_0 + \int{\nabla_x\rho_0(\vec{r}\,)\over
| \vec{r}-\vec{r}\,'|}d\vec{r}\,'\right)_i ~~~~~,~~~~~
Q_{\sigma}(\vec{r}\,)=
{V_{\sigma}\over \rho_0}\nabla_x\rho_0(\vec{r}\,)
\label{aa3}
\end{equation}
with $V_0=V^{xc}(\rho=\rho_0)$.
We have further assumed that  $m_0={2S_z\over N}\rho_0$ in the gs.
Using the results
\begin{eqnarray}
\langle\sum_i Q(\vec{r}_i)\rangle\equiv\int
Q(\vec{r}\,)\delta\rho(\vec{r},t)d\vec{r}=
-\left(\alpha(t)
+ {2S_z\over N}\beta(t)\right)\int \rho_0(\vec{r}\,)
\nabla_x Q(\vec{r}\,)
d\vec{r} \nonumber\\
\langle\sum_i Q_{\sigma}(\vec{r}_i)\sigma_z^i\rangle\equiv\int
Q_{\sigma}(\vec{r}\,)\delta m(\vec{r},t)d\vec{r}=
-\left({2S_z\over N}\alpha(t)
+ \beta(t)\right)\int \rho_0(\vec{r}\,)
\nabla_x Q_{\sigma}(\vec{r}\,) d\vec{r}
\label{aa4}
\end{eqnarray}
it is then possible to write the variations in the one-body
potential of Eq. (\ref{eq1}) in the separable form
\begin{eqnarray}
\delta V(\vec{r}_i,t) = \frac{-1}{1 - (\frac{2S_z}{N})^2}\left[
\left({\langle\sum_i Q(\vec{r}_i)\rangle\over
\int(\nabla_x Q(\vec{r}\,))\rho_0(\vec{r}\,)d\vec{r}} -
{2S_z\over N}{\langle\sum_i
Q_{\sigma}(\vec{r}_i)\sigma_z^i\rangle\over
\int(\nabla_x Q_{\sigma}(\vec{r}\,))\rho_0(\vec{r}\,)d\vec{r}}\right)
\right. \times
\nonumber\\
\left.
\left(Q(\vec{r}_i) +
{2S_z\over N}Q_{\sigma}(\vec{r}_i)\sigma_z^i\right)  +
\left({\langle\sum_i Q_{\sigma}(\vec{r}_i)\sigma_z^i\rangle\over
\int(\nabla_x Q_{\sigma}(\vec{r}\,))\rho_0(\vec{r}\,)d\vec{r}} -
{2S_z\over N}{\langle\sum_i Q(\vec{r}_i)\rangle\over
\int(\nabla_x Q(\vec{r}\,))\rho_0(\vec{r}\,)d\vec{r}} \right)
\right. \times
\nonumber\\
\left.
\left({2S_z\over N}Q(\vec{r}_i) +
Q_{\sigma}(\vec{r}_i)\sigma_z^i\right)\right] ~~.
\label{aa5}
\end{eqnarray}
Following  Ref.\ \onlinecite{Li91}, one could
now express the various responses
 to an oscillating potential of the form (\ref{aa5})
in terms
of the independent particle response functions $\chi^0_{\sigma,\sigma'}$
through RPA-type equations. However, our aim here is to develop an
analytical model which allows us to understand in a simple way the
numerical results of Section III. To this end we take in
(\ref{aa3})
a harmonic oscillator for the one-particle potential
$V_0={1\over2}\omega_{xc}^2r^2$ to simulate the short range
effects, and a step
function for the electronic density $\rho_0$ entering the long range
contribution $\int \nabla_x\rho_0(\vec{r}\,)/
|\vec{r}-\vec{r}\,'|d\vec{r}\,'$. We also assume
$Q_{\sigma}(\vec{r}_i)  \simeq -2 V_{\sigma} x_i / \langle r^2\rangle$.
Equation (\ref{eq16}) of Section IV is then obtained.

\begin{figure}
\caption{Dipole strength function (effective atomic units) of the $N$
= 5 dot as a function of frequency  (meV). Solid and dotted
lines correspond to the density response to $D_{\rho}$,
and to the spin  response to $D_m$, respectively. Dashed lines represent
the free particle strength function.}
\label{fig1}
\end{figure}
\begin{figure}
\caption{Same as Fig. 1 for $N$ = 25.}
\label{fig2}
\end{figure}
\begin{figure}
\caption{Same as Fig. 1 for $N$ = 210.}
\label{fig3}
\end{figure}
\begin{figure}
\caption{Same as Fig. 2 using a parabolic potential with
$\omega_0$  = 2.78 meV instead of a disk confining potential.}
\label{fig4}
\end{figure}
\begin{figure}
\caption{Mixed $S_{m\rho}(\omega)$ response function
 (effective atomic units)  of the $N$ = 210
dot at $\nu$ = 3. The arrows indicate the density and spin mode peaks.}
\label{fig5}
\end{figure}
\begin{figure}
\caption{$B$ dispersion of the main peaks of the $N$ = 5 spectrum.
The circles correspond to density modes and the triangles to spin
modes. The solid symbols correspond to $\nu$ = 1.
The lines represent the VPM $B$ dispersion laws with
fitted value at $B$ = 0.}
\label{fig6}
\end{figure}
\begin{figure}
\caption{$B$ dispersion of the main peaks of the
$N$ = 25 spectrum for a disk confining potential.
The circles correspond to density modes and the triangles to spin
modes. The crosses are experimental points from Ref.[2].
The lines represent the VPM $B$ dispersion laws with
fitted values at $B$ = 0.
The insert shows the negative $B$ dispersion branch
of the spin mode. From left to right the solid symbols correspond to
$\nu$ = 6 to 1.}
\label{fig7}
\end{figure}
\begin{figure}
\caption{$B$ dispersion of the main peaks of the
$N$ = 210 spectrum for a disk confining potential.
The circles correspond to density modes and the triangles to spin
modes. The crosses are experimental points from Ref.[2].
The lines represent the VPM $B$ dispersion laws with
fitted values at $B$ = 0.
The insert shows the negative $B$ dispersion branch
of the spin mode. From left to right the solid symbols correspond to
$\nu$ = 8 to 1.}
\label{fig8}
\end{figure}
\begin{figure}
\caption{Same as Fig. 7 using a parabolic confining potential with
$\omega_0$  = 2.78 meV instead of a disk confining potential.}
\label{fig9}
\end{figure}
\begin{figure}
\caption{Dipole responses for the $N$ = 210 dot at $B$ = 7 T.
Top panel, $T$ = 0.1 K. Bottom panel, $T$ = 2 K. Lines as indicated
in Fig. 1.}
\label{fig9b}
\end{figure}
\begin{figure}
\caption{$B$ dispersion of the spin dipole mode within  VPM.
Experimental points are from Ref. [13].}
\label{fig10}
\end{figure}

\begin{references}
\bibitem{note} The results at $T$ = 2 K displayed in Fig.
\ref{fig9b} have been obtained with
an imaginary part of the frequency $\eta$ = 0.015 instead of the 0.006
value used everywere else,  to make sure that the condition
$\eta > \Delta \omega > k_B T $ is fulfilled.
\end{references}
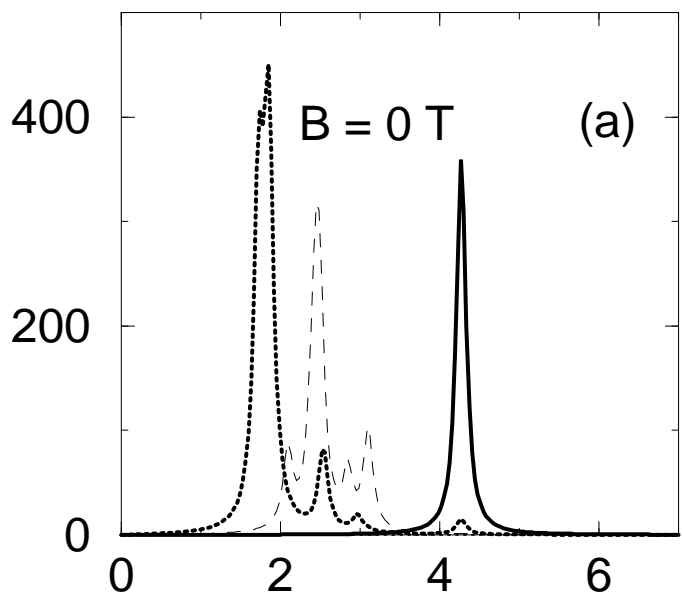
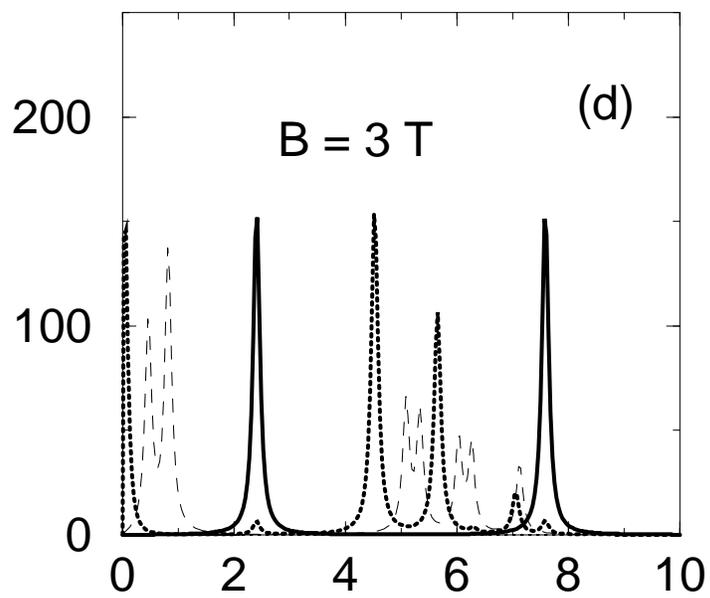
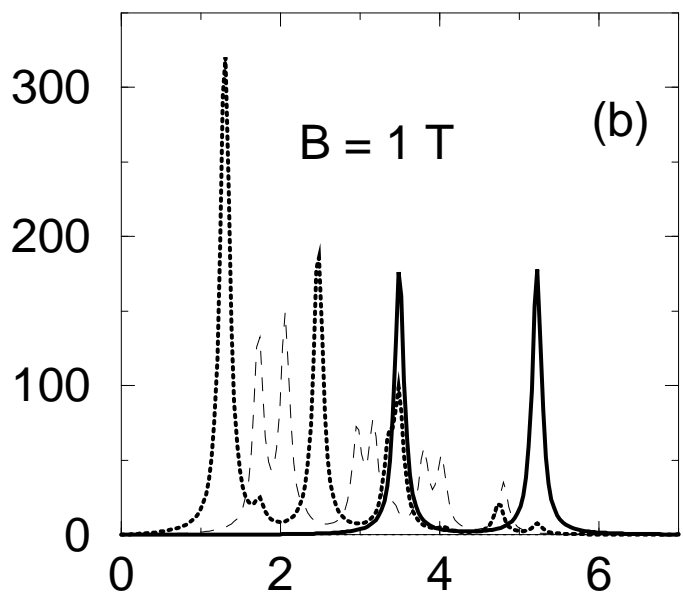
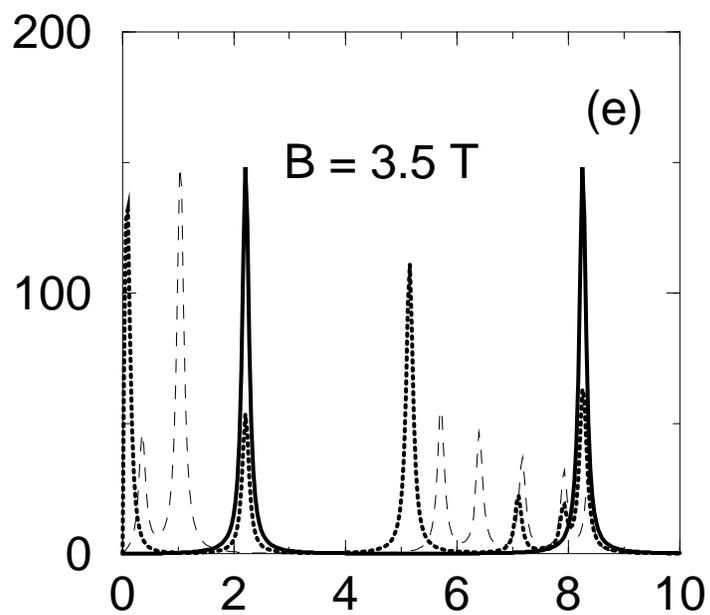
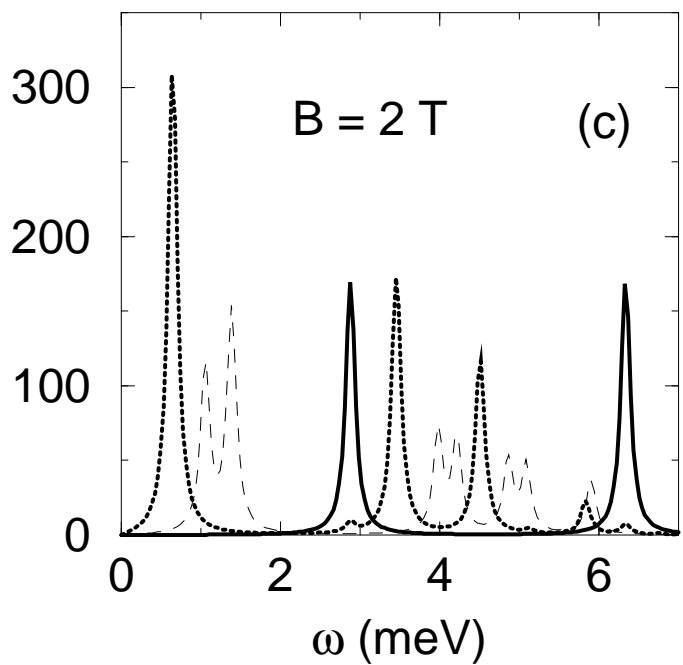
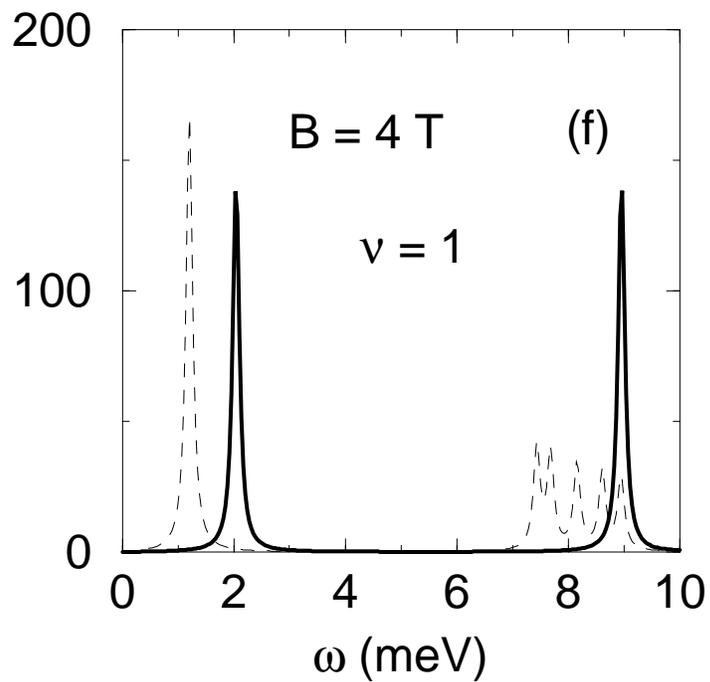
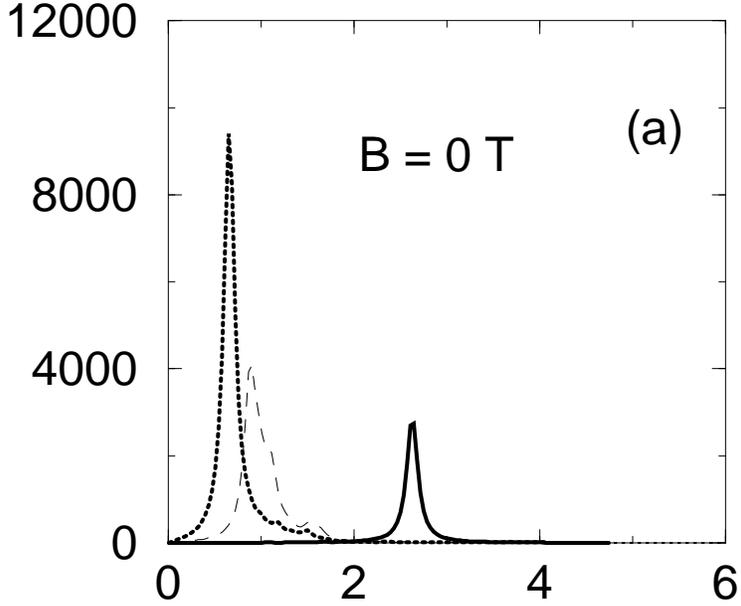
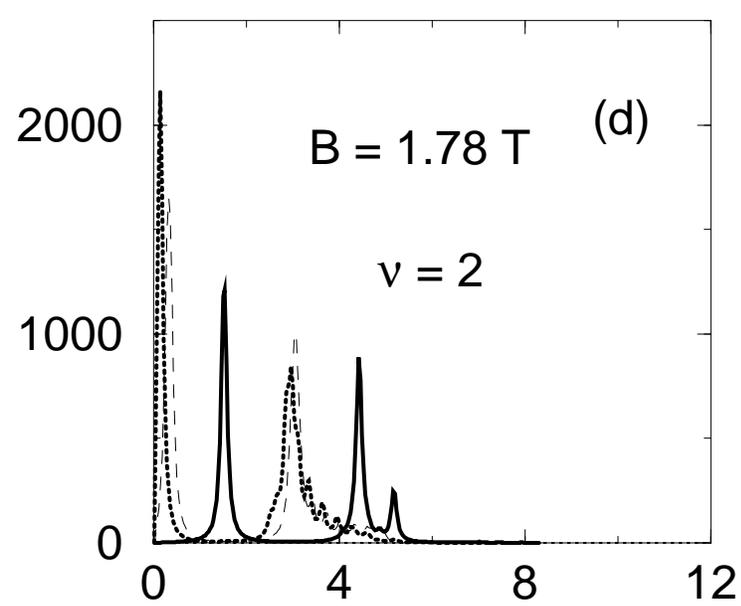
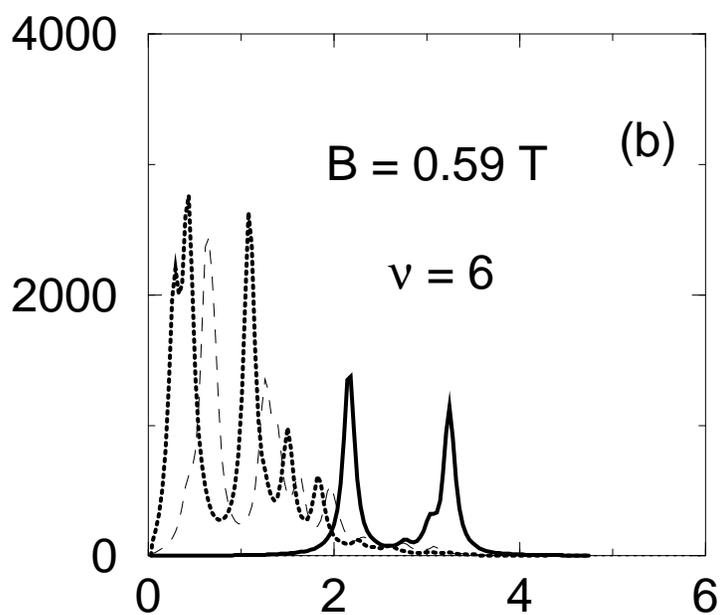
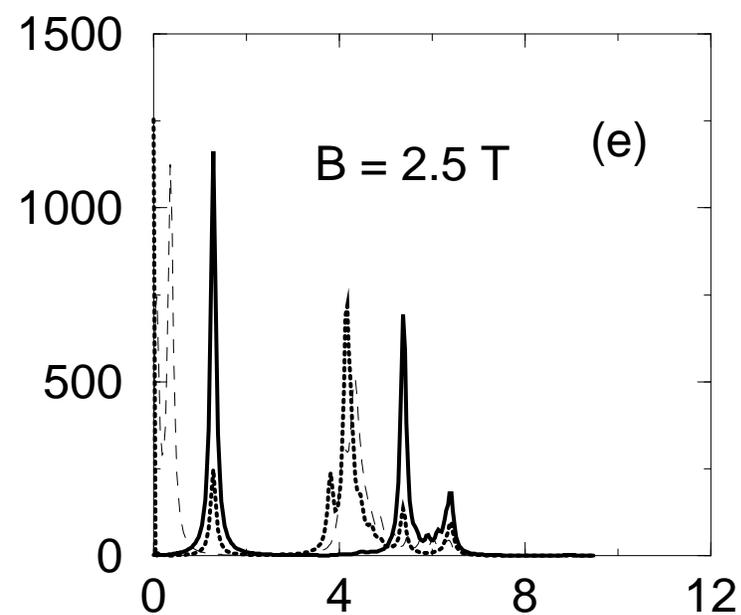
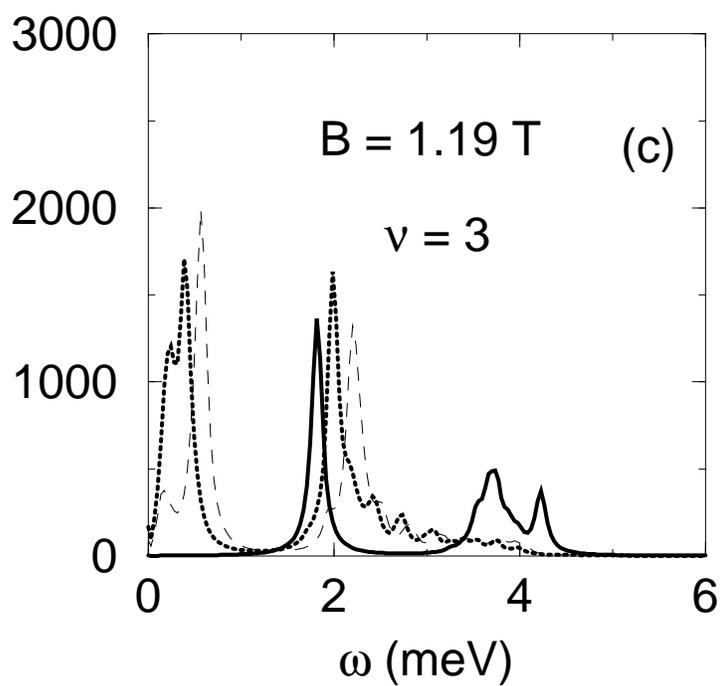
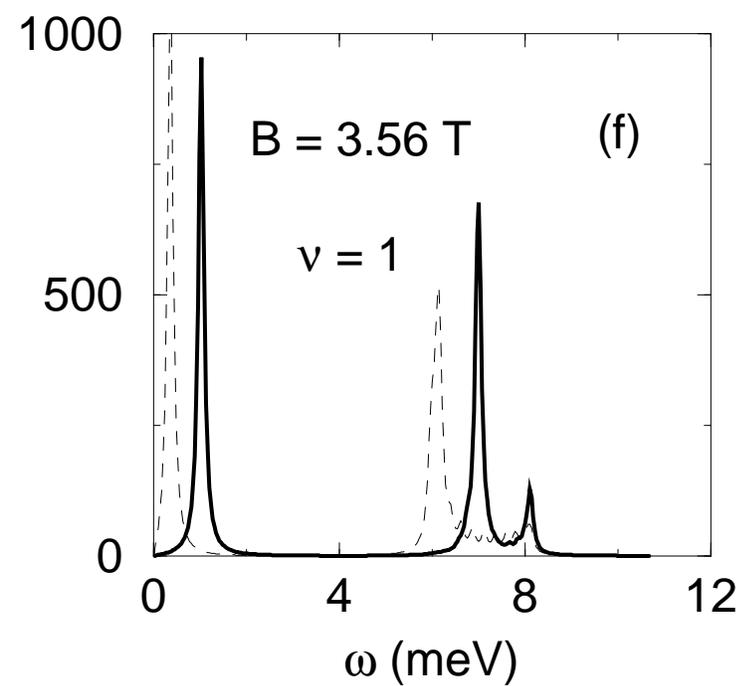
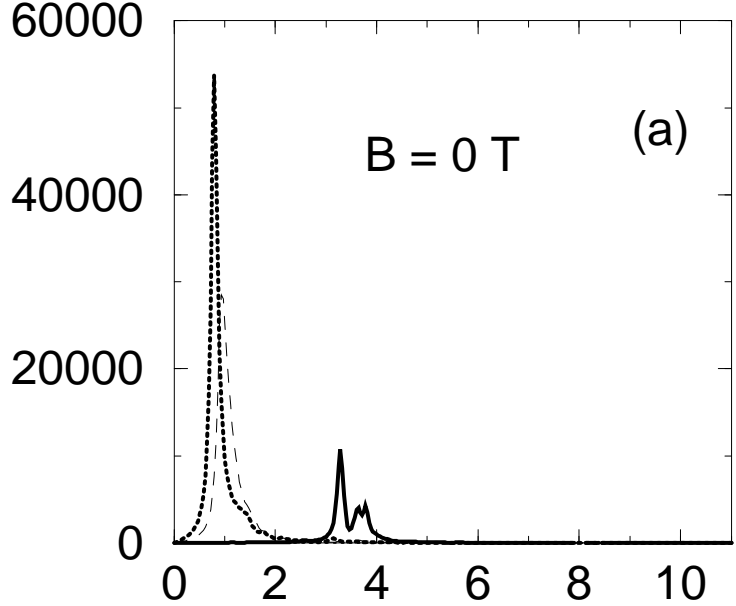
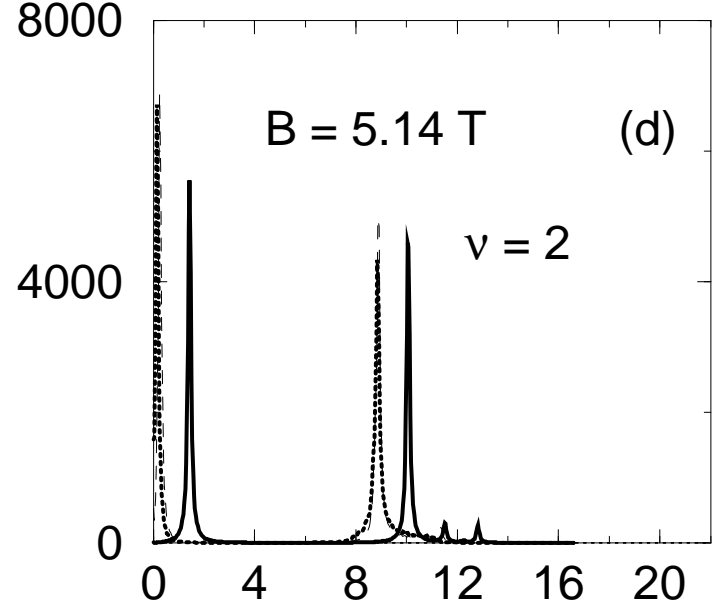
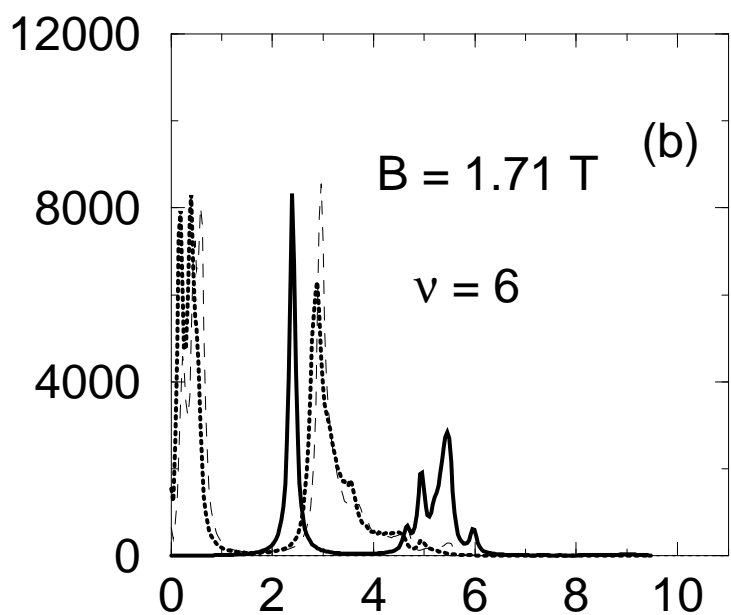
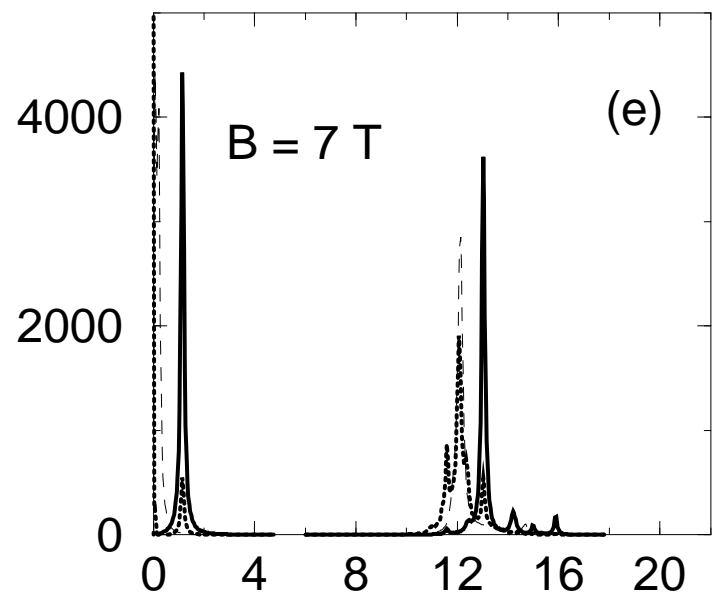
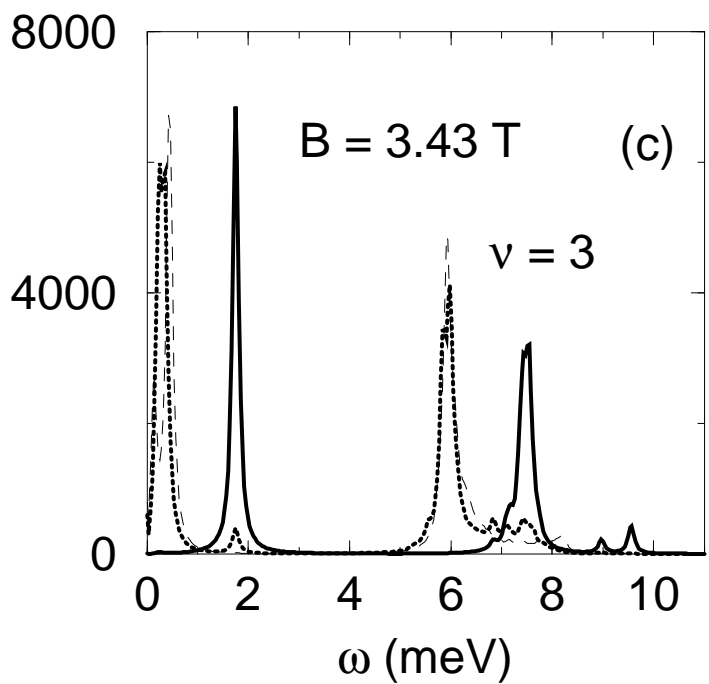
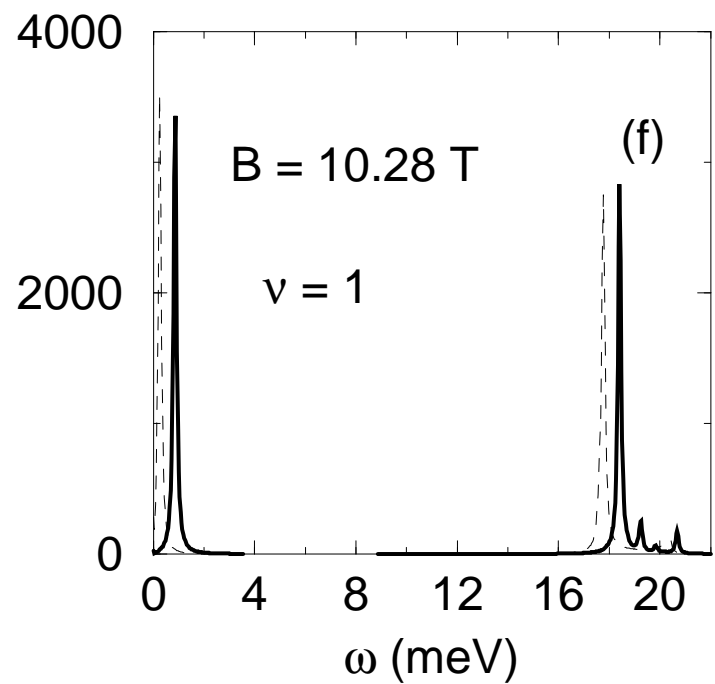
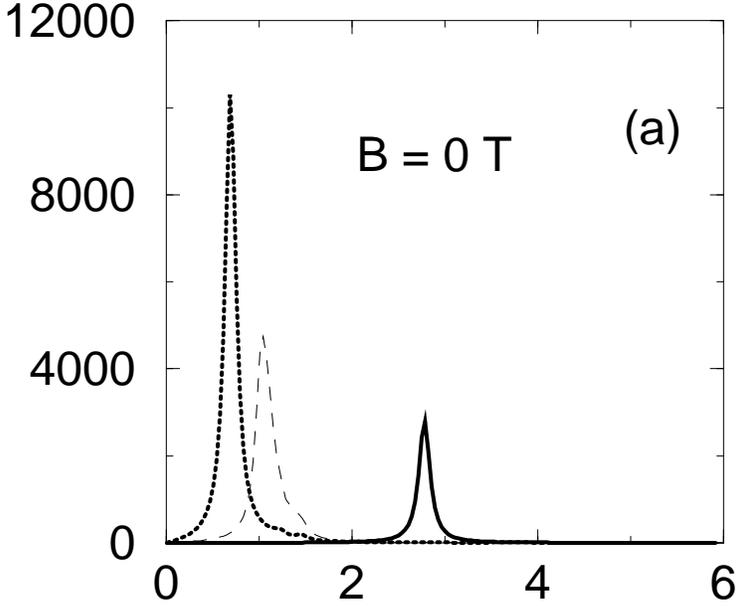
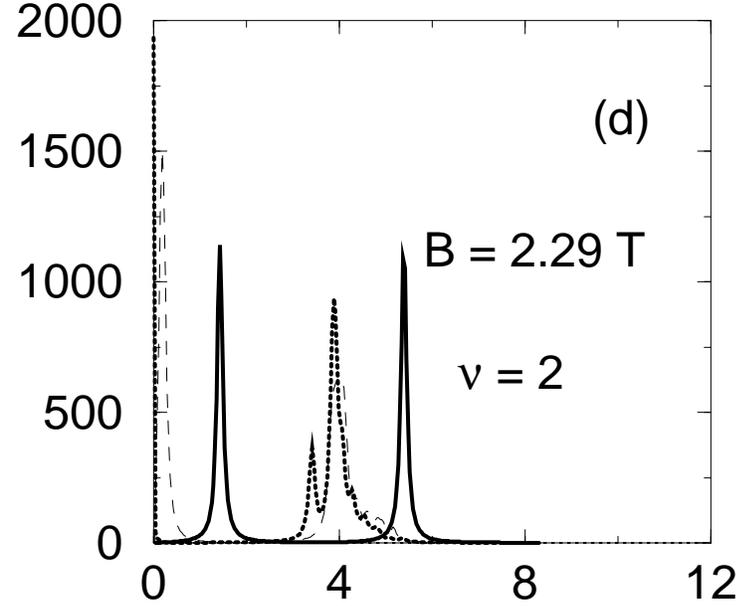
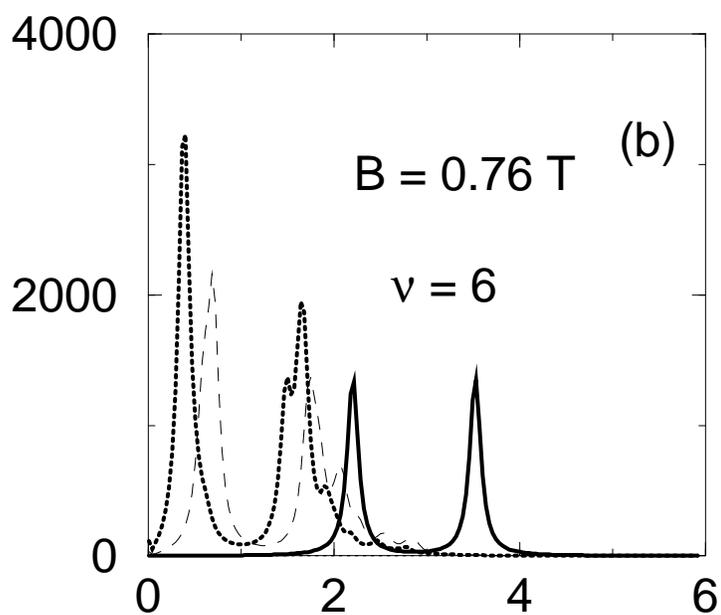
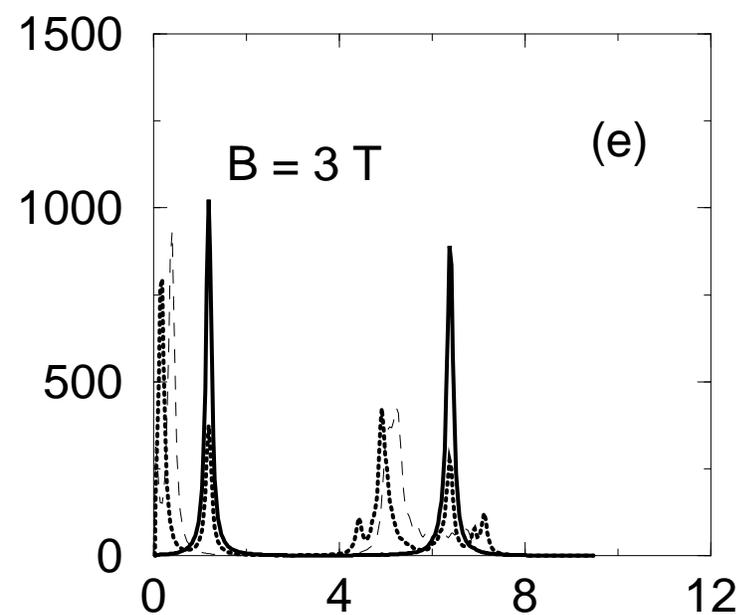
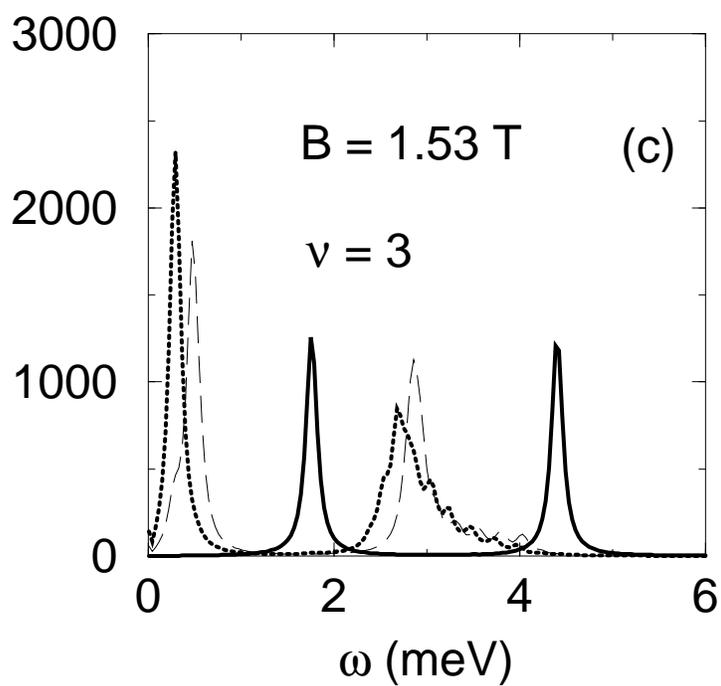
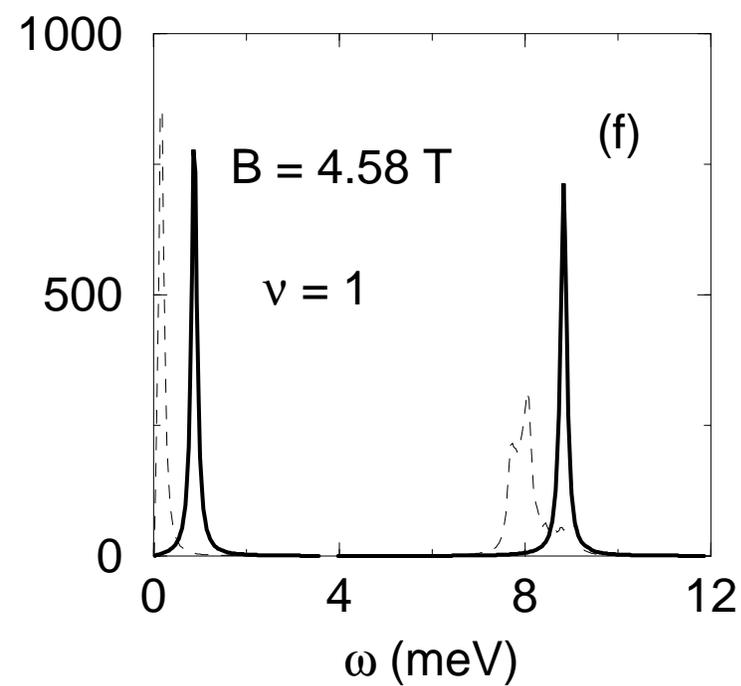
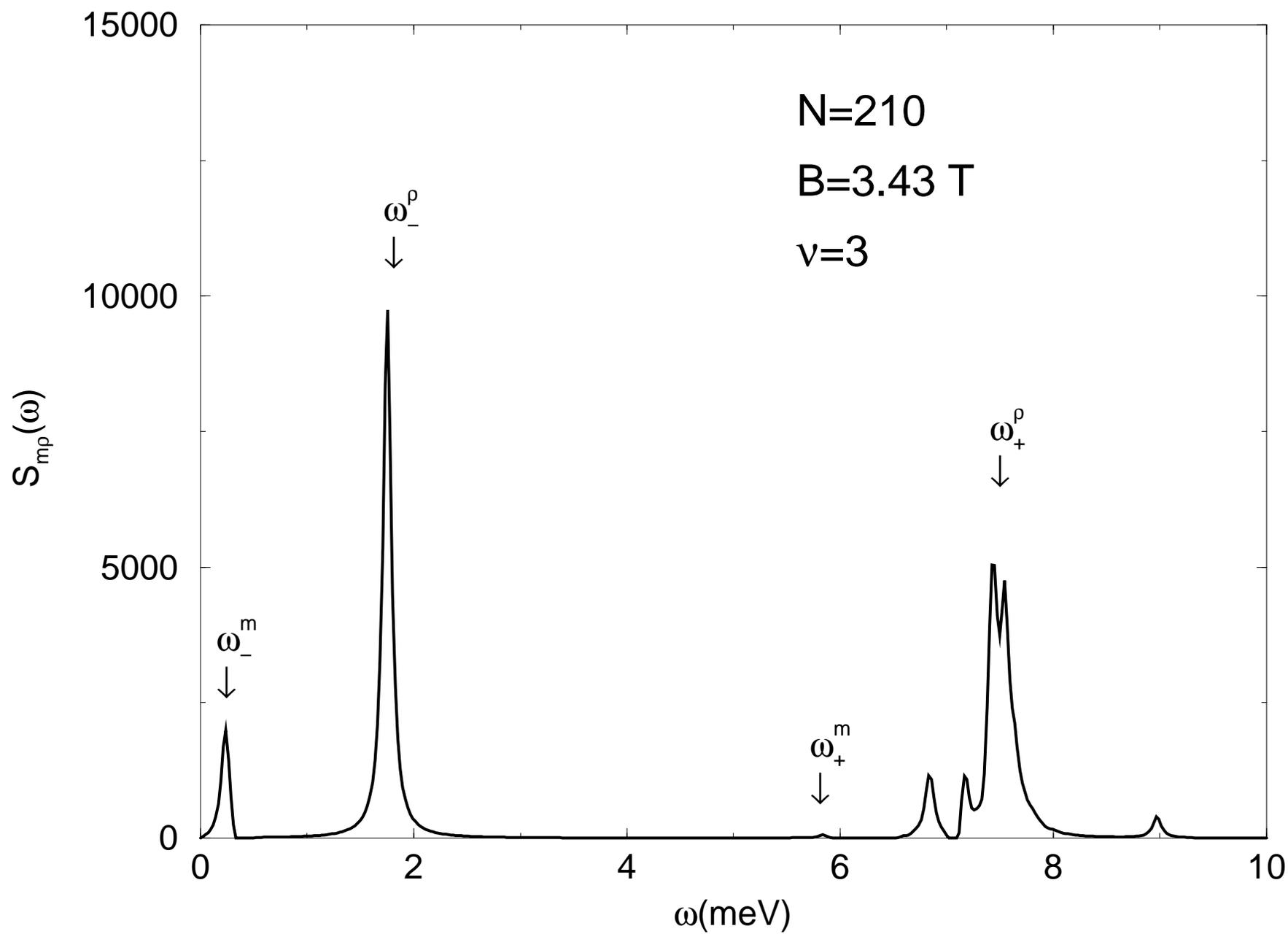
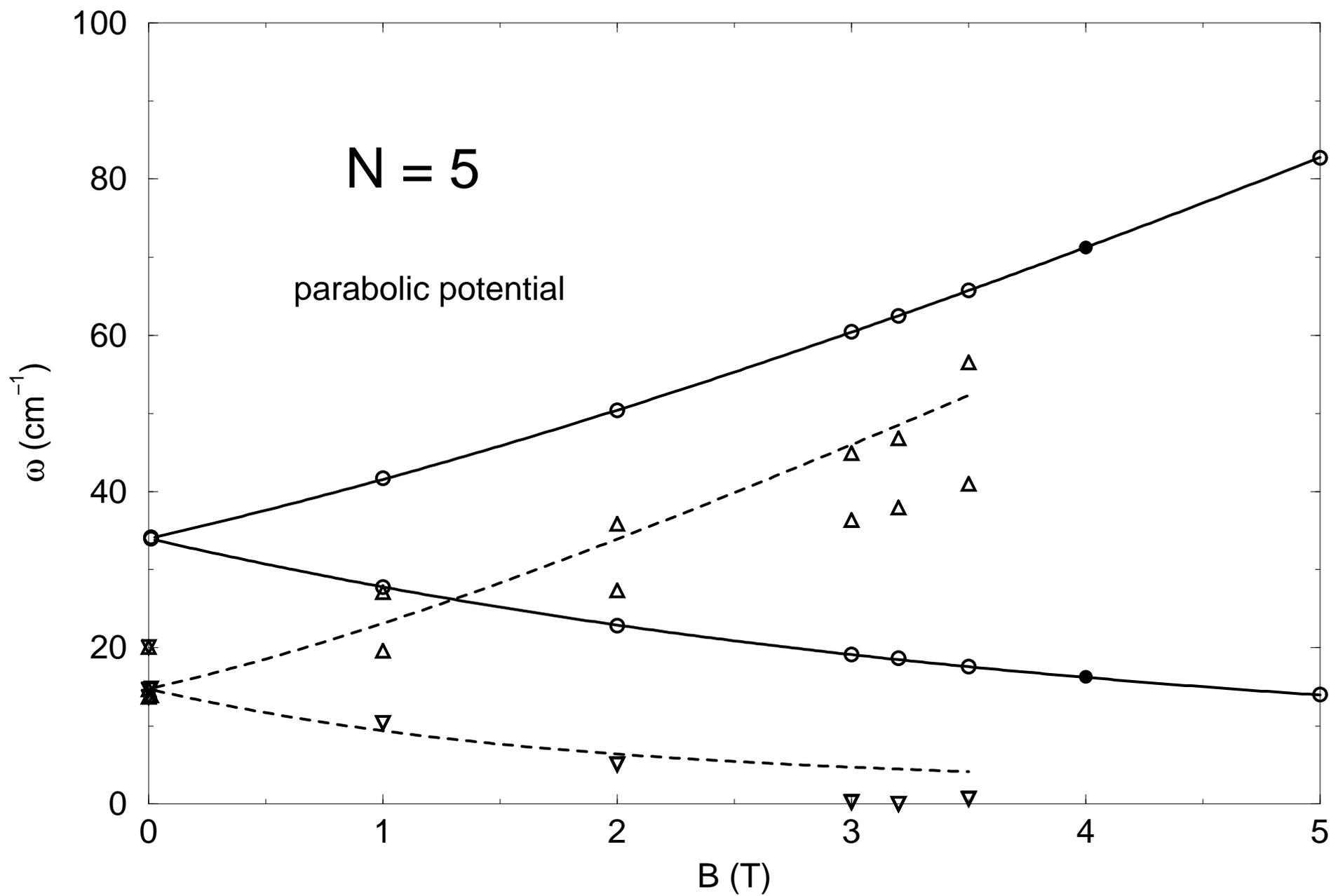
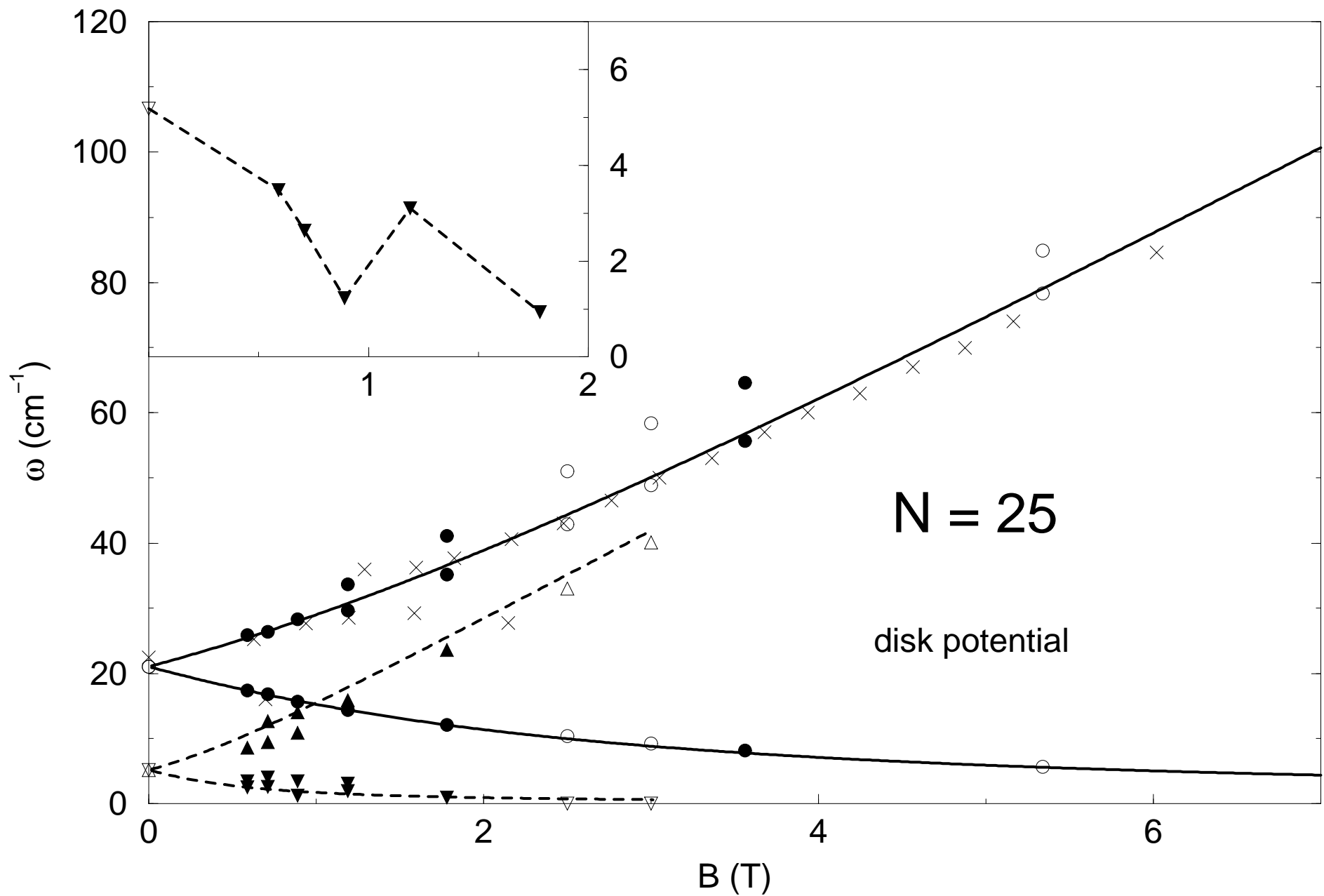
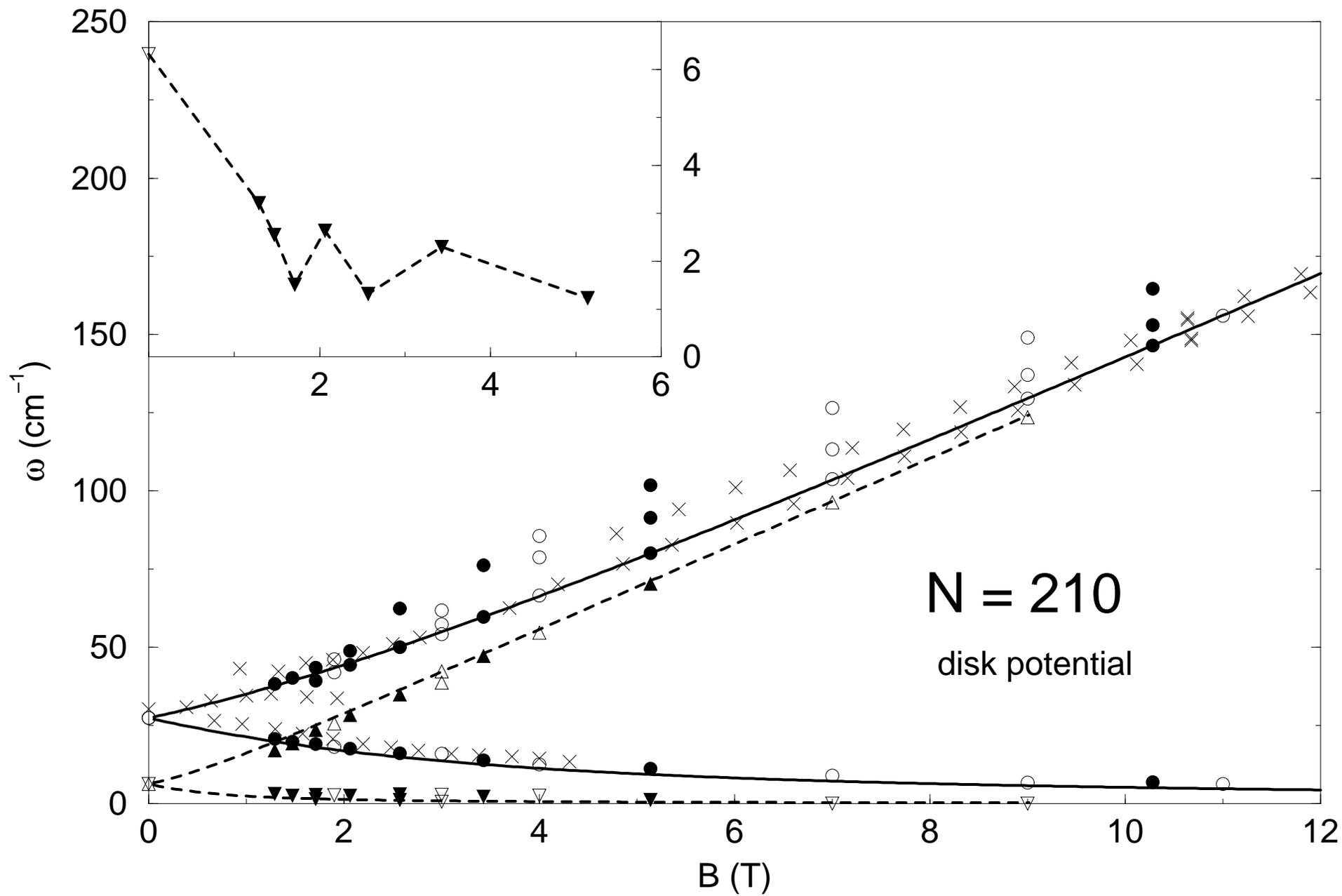
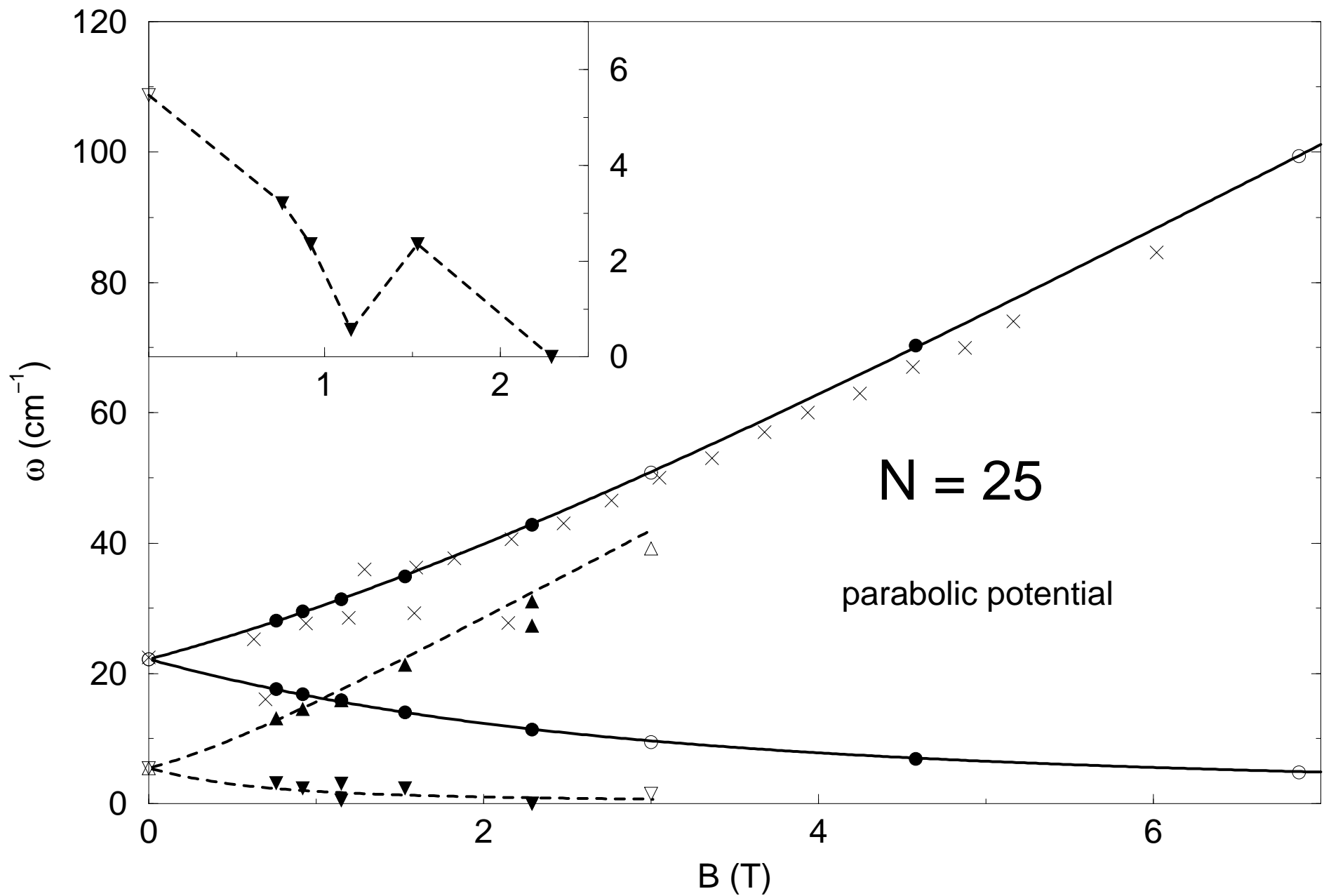
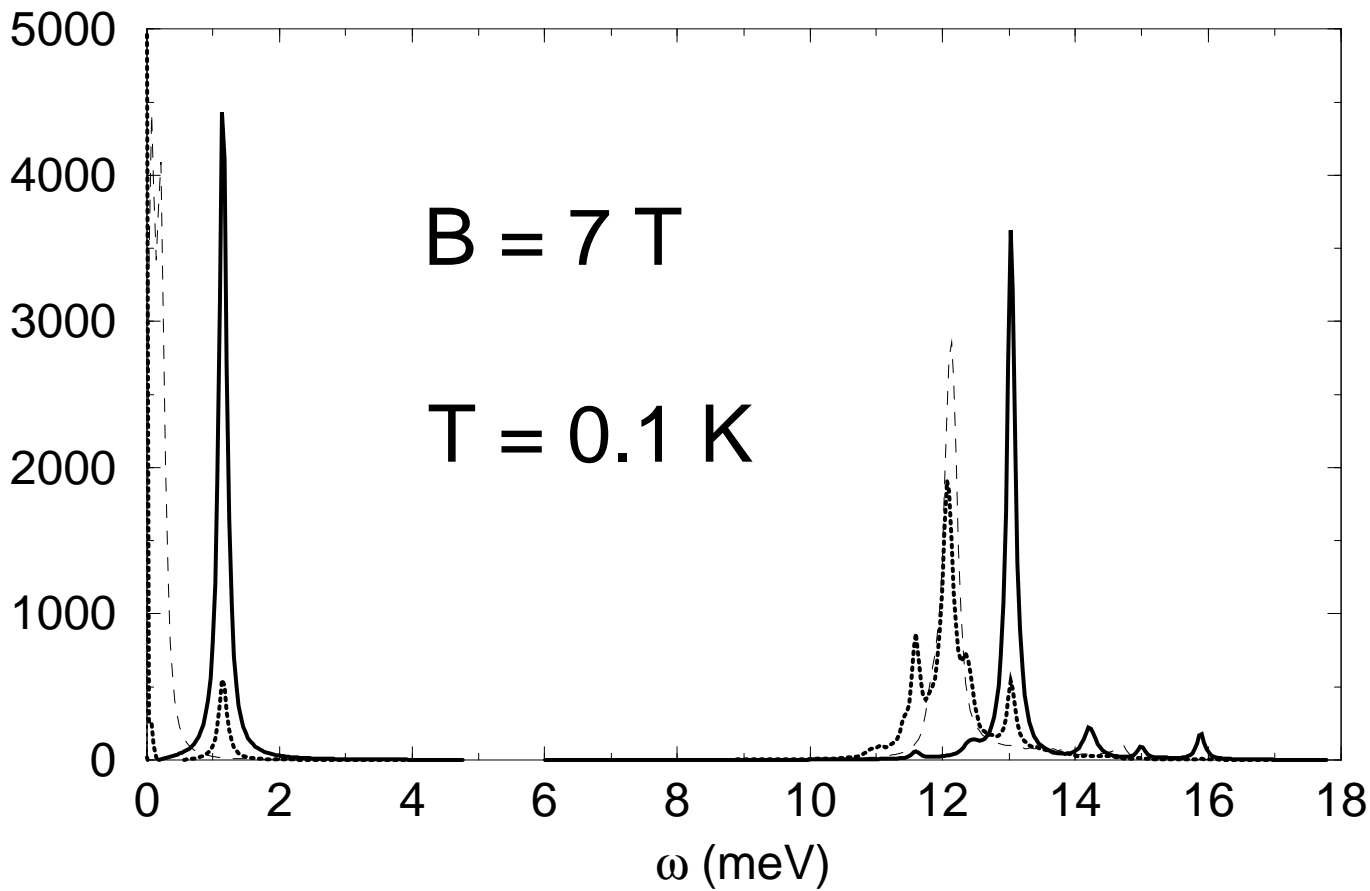
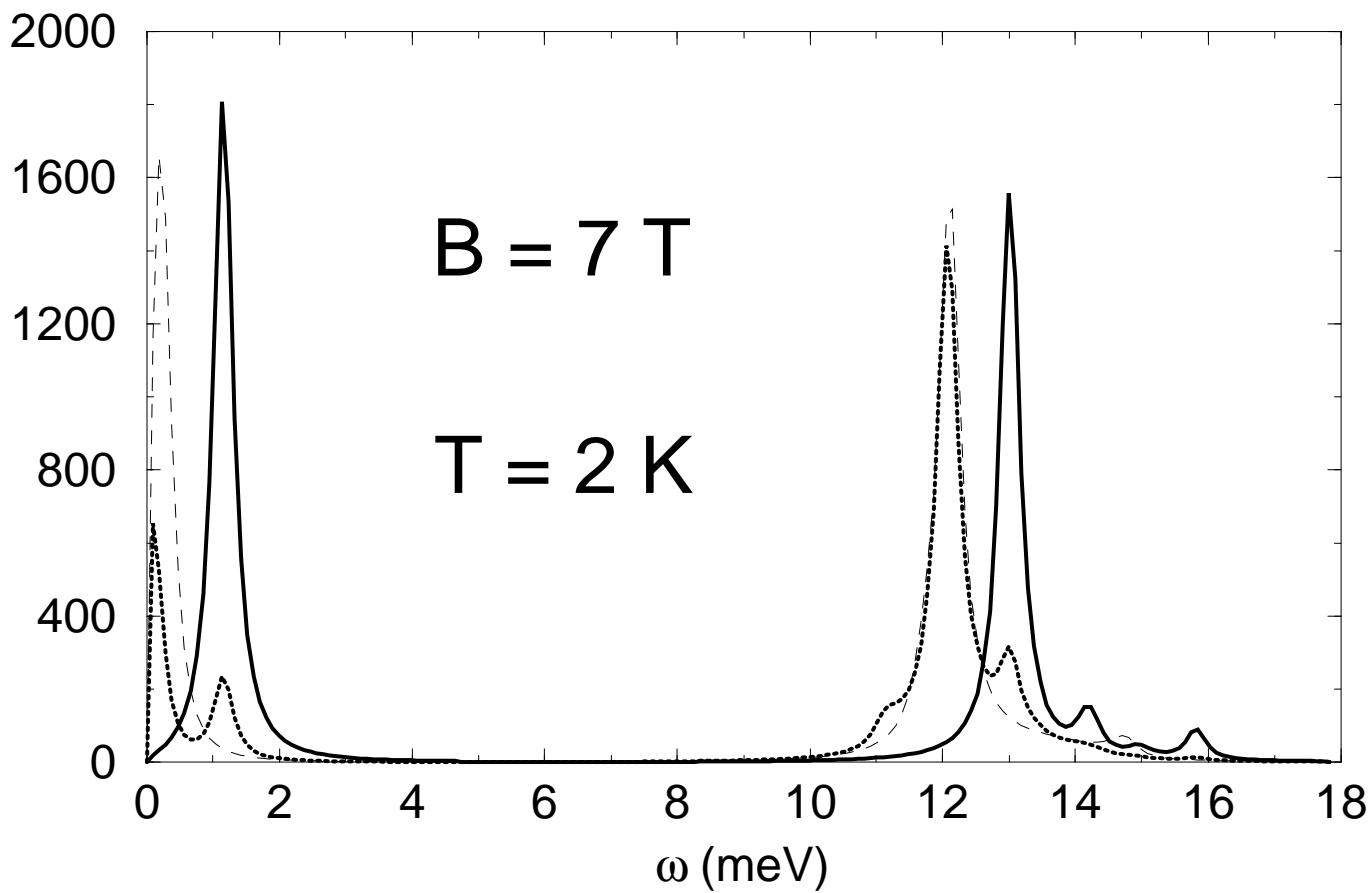
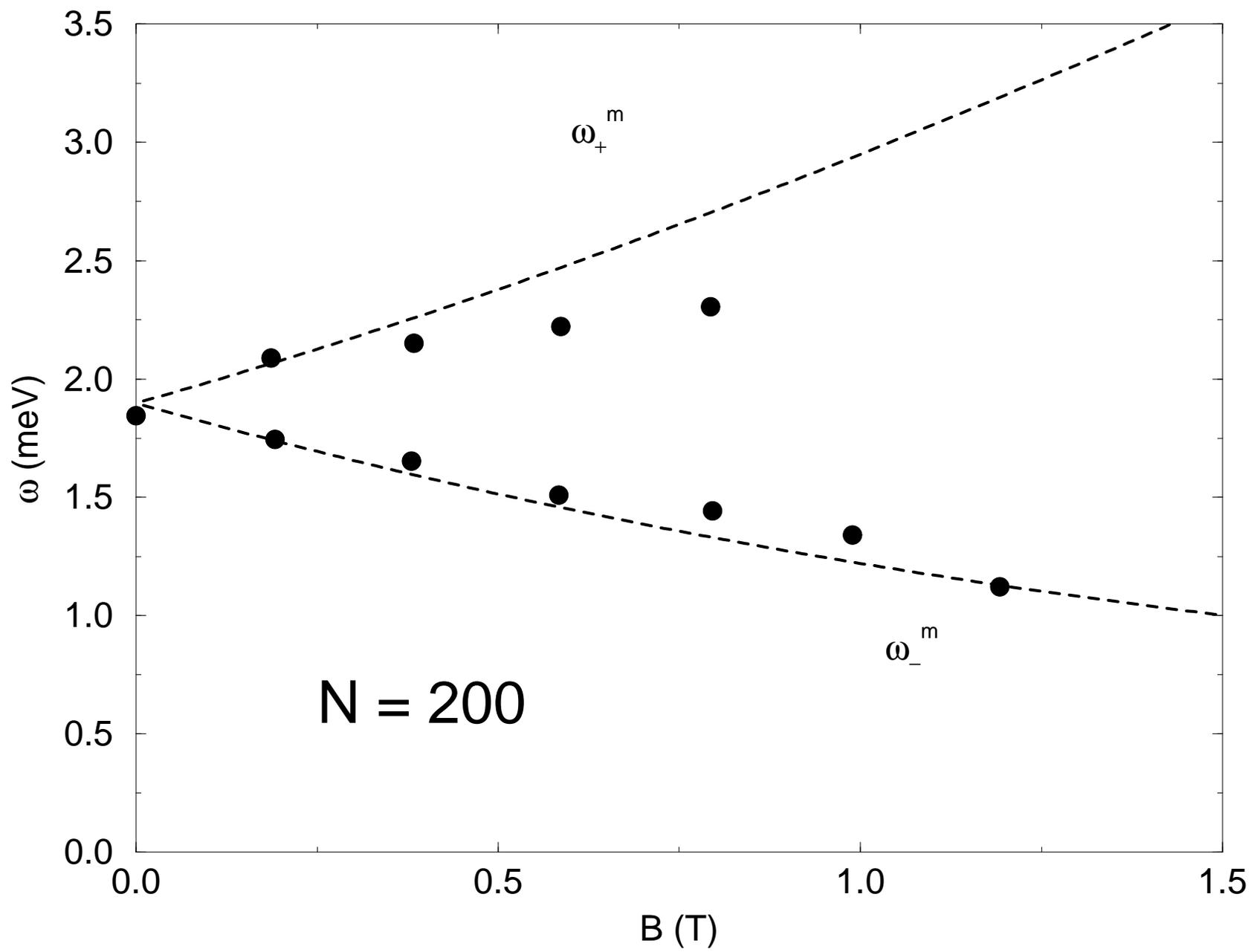
\end{document}